\documentclass[final,12pt]{article}
\usepackage[usenames,dvipdfmx]{color}
\usepackage{setspace,amsfonts,comment,amsmath,amsthm,amssymb,amsxtra,graphicx}
\usepackage[pdftex,pagebackref]{hyperref}
\usepackage{mathpazo}

\usepackage{threeparttable}
\usepackage{booktabs}
\usepackage[T1]{fontenc}
\usepackage{inputenc}  %[latin9]
\usepackage{esint}

\usepackage[authoryear]{natbib}
\usepackage{sidecap}
\usepackage[font=footnotesize]{caption}
\usepackage{pgfplots}
\usepackage{tabularx}% http://ctan.org/pkg/tabularx
\usepackage{booktabs}% http://ctan.org/pkg/booktabs
\pgfplotsset{width=7cm,compat=1.8}

\usepackage{float}
\hypersetup{pdfpagemode=None, colorlinks=true,
 anchorcolor= webbrown, citecolor= blue,
filecolor= webbrown, linkcolor= blue, menucolor= webbrown,
urlcolor= webbrown, citebordercolor= 1 0 0, menubordercolor=1 0
0, urlbordercolor=1 0 0, runbordercolor=1
0 0 } \hypersetup{pdfauthor=Bruno Levy}
\definecolor{webgreen}{rgb}{.6,.6,.6}
\definecolor{webbrown}{rgb}{.6,0.15,0.15}
\definecolor{webyellow}{rgb}{0.98,0.92,0.73}

\newcommand\ddate{\today}
\newcommand\aauthorA{\footnote{brunopcl@al.insper.edu.br} Bruno P. C. Levy  }
\newcommand\aauthorB{\footnote{hedibertfl@insper.edu.br} Hedibert F. Lopes}
\newcommand\aaffiliationA{Insper}
\newcommand\aaffiliationB{Insper}
\newcommand\eemailA{\href{mailto:brunopcl@al.insper.edu.br}{{\small\texttt{{}}}}}
\newcommand\eemailB{\href{mailto:hedibertfl@insper.edu.br}{{\small\texttt{{}}}}}
\newcommand\ttitle{Trend-Following strategies via Dynamic Momentum Learning} %Time Series Momentum Predictability via Dynamic Binary Classification
\newcommand\tthanks{We thank to all participants at Insper Seminars for useful comments. All remaining errors are of our responsibility.
}

\newcommand{\propref}[1]{\hyperref[#1]{Proposition~\ref*{#1}}}
\newcommand{\appref}[1]{\hyperref[#1]{Appendix~\ref*{#1}}}
\newcommand{\secref}[1]{\hyperref[#1]{Section~\ref*{#1}}}

\newcommand{\commentout}[1]{}

\newtheorem*{lemma*}{Lemma}

\hypersetup{pdfstartview=FitH}
\title{\vspace{-1.5\baselineskip}  % \Huge
{\ttitle\thanks{~\tthanks}}\bigskip}
\author{{\aauthorA} \\   % \textbf
{\small \aaffiliationA}\\
[-.1in] \eemailA \and
{\aauthorB} \\       % \textbf
{\small \aaffiliationB}\\
[-.1in] \eemailB \medskip }
\date{\normalsize \textcolor{webgreen}{\ddate}}
\date{\normalsize \textcolor{webgreen}{Working Paper - September, 2021}}

\begin{document}

\maketitle

\vspace{-1.5\baselineskip}

\begin{abstract} 

Time series momentum strategies are widely applied in the quantitative financial industry and its academic research has grown rapidly since the work of \cite*{moskowitz2012time}. However, trading signals are usually obtained via simple observation of past return measurements. In this article we study the benefits of incorporating dynamic econometric models to sequentially learn the time-varying importance of different look-back periods for individual assets. By the use of a dynamic binary classifier model, the investor is able to switch between time-varying or constant relations between past momentum and future returns, dynamically combining or selecting different momentum speeds during turning points, improving trading signals accuracy and portfolio performance. Using data from 56 future contracts we show that a mean-variance investor will be willing to pay a considerable management fee to switch from the traditional naive time series momentum strategy to the dynamic classifier approach.

\noindent \textbf{Keywords: Time-Series Momentum; Dynamic Classifier; Dynamic Portfolio Allocation; Crashes; Binary Forecasting}.

\noindent\noindent \emph{J.E.L. codes: C11, C53, C58, G11, G17 }

\end{abstract}

\onehalfspace

\newpage

\tableofcontents

\newpage

\section{Introduction} \label{sec1: Introduction}

A significant part of the hedge fund industry nowadays is based on managed futures funds, also known as Commodity Trading Advisors (CTAs). As shown by \cite*{hurst2013demystifying}, the returns of these funds are usually explained by simple trend-following (aka time-series momentum)  strategies on future contracts. These strategies use the ability of past returns to antecipate future return movements. The work of \cite*{moskowitz2012time} was the first to document the ability of time-series momentum strategies to generate significant profits over time and among different future markets, contradicting the random-walk theory where no past information is able to predict future returns. The basic ideia of such strategy is to vary the position of an individual asset based on signals of the past returns over a specific look-back period (traditionally, from one to twelve months). Therefore, the investor goes long during periods of positive trends and goes short during periods of downtrend.

The time-series momentum strategy is related to, but different from, the cross-sectional momentum strategy (\citealp{jegadeesh1993returns} and \citealp{asness2013value}). The cross-sectional approach explores the \textit{relative} performance among different assets, buying those assets with higher past performance (winners) and selling those with lower performance (losers). Hence, even a security with positive but low past return can be sold if its peers are performing better recently. On the other hand, the time-series momentum explores the \textit{absolute} performance of the own specific security, despite the performance of its peers. Interestingly, the work of \citealp{moskowitz2012time} shows that the returns of time-series momentum strategies are not related to compensation for traditional risk factors, such as the value and size factors, but is partially related to the momentum factor.

After the work of \citealp{moskowitz2012time},  the  empirical  literature  on  time-series momentum has  grown  rapidly, finding evidences that the returns of managed funds can be explained by time-series momentum strategies (\citealp{hurst2013demystifying} and \citealp{baltas2013momentum}) and its significant performance in different asset classes in emerging and developed markets (\citealp{georgopoulou2017trend}), among common stocks (\citealp*{lim2018time}) and throughout the entire past century (\citealp{hurst2017century}).
Using intraday data, \cite*{gao2018market} also show that the first half-hour return on the market is able to predict the last half-hour return.  In terms of portfolio allocation, \cite{baltas2015trend}, \cite{baltas2020demystifying}
 and \cite{rubesam2020long} show the benefits of correlations and risk parity for improving portfolio diversification on time-series momentum strategies.

Recently, \cite{hutchinson2020time} have showed a link between time-series momentum returns and the business cycle, giving evidences that the returns are stronger during both recessions and expansions. The literature has also recognized the time-series momentum pattern in risk factors. \cite{gupta2019factor} document robust persistence in the returns of equity factor portfolios, showing that factor timing by time-series momentum produces economically and statistically large
excess performance relative to untimed factors. Exploring this idea, \cite{levy2021dynamic} also insert a time-series momentum structure to predict risk factors in a high-dimensional portfolio allocation.

In general, the papers cited above compare the results of different portfolios built by the use of different look-back periods (the number of periods to consider in the past to form a measure of momentum) or directly consider twelve months as the benchmark measure to generate momentum signs. Then, they set a buy or sell trading rule based on the observed momentum sign. This type of decision rule is motivated by practice and the academic literature that followed. However, we argue in this paper that the absence of an econometric model behind decisions can lead investor to misleading trading actions. For example, what guarantees that the returns from previous months will always indicate a positive relationship with future returns? Each asset can respond differently not just to the same measure of momentum but also for different look-back periods. Some assets can have a negative (reversal) relation with shorter look-back periods and others a positive effect. Also, this pattern can change over time. Since the environment of the financial market is continuously changing, a pattern that was common in the 80s can differ from the 90s or during financial crisis and pandemics. Motivated by these ideas, we use a dynamic binary classification model to infer about the future trend of returns. The approach is able to handle look-back period uncertainty and time-varying parameters in a dynamic fashion. Hence, investors can learn from past mistakes, giving lower importance to look-back periods that have performed worse in the recent past and assigning higher probabilities to look-back periods with higher predictability. Also, by the use of time-varying parameters, the model adapts to changes in the financial environment, switching from periods of momentum to reversal if it is empirically wanted.

The literature on return predictability is not new. The seminal paper of \cite{welch2008comprehensive} shows that it is extremely hard to predict stock returns using well known predictors in a econometric model, i.e., predictors are not able to outperform the simple historical average of stock returns. After \cite{welch2008comprehensive}, several other studies have appeared in the literature trying to find bettter predictors or econometric models that could be able to improve predictability (\citealp{campbell2008predicting}, \citealp*{rapach2010out}, \citealp{dangl2012predictive}, \citealp*{johannes2014sequential}, \citealp*{chinco2019sparse}, \citealp*{gu2020empirical}, \citealp*{liu2021can} and many others). Some crucial aspects that can be found in several papers that followed \cite{welch2008comprehensive} are the presence of time-varying coefficients and model combination. In fact, the accumulated academic evidence has shown that parameter instability is able to handle changes in market sentiment, institutional framework and macroeconomic conditions. Additionally, model combination is able to dramatically improve forecasts since it combines important economic information contained in each different predictor.    

Inspired by the recent advances on the return predictability literature, our goal is to improve trend-following strategies by the use of model selection and model combination, where different look-back periods can be considered to build momentum measures. We follow the approach of \cite*{mccormick2012dynamic} to build our dynamic trend return classifier. Our classifier relies on the use of a dynamic logistic regression where parameters are able to be constant or time-varying over time and uncertainties about how far the investor should look into the past to predict the future is dealt by the use of dynamic model probabilities. After assigning probabilities for each model setting, we are able to integrate uncertainties by dynamic model averaging (DMA) or dynamic model selection (DMS). The approach is a binary counterpart of the DMA approach recently used with great sucess in other Bayesian econometric applications (\citealp{koop2012forecasting}, \citealp{dangl2012predictive}, \citealp{koop2013large}, \citealp*{catania2019forecasting} and \citealp{levy2021dynamic_b}). Using discounting methods and distribution approximations, there is no need to use expensive simulation schemes such as Markov Chain Monte Carlos (MCMC), which makes the whole process much faster to compute. It can be viewed as a great advantage for quantitative investors, since the amount of assets available is growing and trading positions are getting faster nowadays.

The binary approach of \cite{mccormick2012dynamic} was originally applied to a medical classification problem and it was first introduced in the economic literature in  \cite{hwang2019forecasting} where the authors use the binary classification method to forecast recession periods. At the best of our knowledge, we are the first to introduce this dynamic approach in a financial econometric context. Since our interest here is not to predict raw returns but its future direction (buy or sell sign), it fits perfectly to the time-series momentum application. The great advantage of using dynamic model probabilities is to combine different economic informations coming from many look-back periods in a sequential fashion. As soon as new data arrives, the model is able to adapt to new informations, assigning higher probabilities for models using look-back periods with stronger informations.

The idea of combining information from different look-back periods has already appeared in the literature before. \cite*{han2016trend} show economic gains when combining informations from short, intermediate and long-term look-back periods to build cross-sectional momentum strategies. More recently, the works of \cite*{garg2020breaking} and \cite*{garg2021momentum} explore the impacts of turning points on time series momentum strategies. They show evidences of an increase in the presence of trend breaks in the last decade, leading to a negative impact on final portfolio performance. It happens due to the fact that after a trend reverses its direction, trend-following strategies tend to place bad bets since past momentum can reflect an old and inactive trend direction. The authors propose a trading rule where information of both fast and slow momentum look-back periods are considered if it identifies a turning point. Also, by the use of a machine learning technique, the  work of \cite*{jiang2020re} use stock-level prices images to detect future price directions instead of using returns information. They apply a convolutional neural network model to classify future return signals and perform a cross-sectional portfolio strategy based on these signal predictions. The authors found robust evidences that image-based predictions are powerful to predict future returns.

Additionally to the increase in trend breaks in the last decade, in Section \ref{crash_2009} we also discuss a topic not well explored by the academic literature on time series momentum: the impacts of the 2009 market rebound on portfolio performance and drawdowns. Similar to the momentum crash observed on cross-sectional momentum strategies after the Great Financial Crisis (\citealp{barroso2015momentum} and \citealp{daniel2016momentum}), traditional time series momentum portfolios also suffered from strong trend breaks, leading to huge losses as soon as old negative trends reverted to positive ones. Motivated by the literature on time-series momentum and return predictability, our goal is to provide an econometric solution to deal with trend reversals, minimizing portfolio drawdowns.

The great advantage of our classifier model compared to the works mentioned above is its ability to sequentially learn the importance of each look-back period individually and for each asset in parallel. Using a dynamic model, we are able to understand the time-varying behavior among different momentum speeds individually and assign higher or lower speed probabilities which are updated from most recent data observations. Hence, as soon as a market correction or rebound seems to appear in the data, slower momentum measures start to receive lower probabilities while faster momentum probabilities increase, influencing final predictions. Therefore, the dynamic classifier approach is able to deal with turning points problems in a customizable and automatic fashion. Also, by allowing time-varying parameters, the model introduces higher flexibility to capture positive or negative relations among past accumulated returns and future returns. Hence, for some periods of time, past returns can induce reversal but in others, momentum.

Using futures data from 1980 to September 2020 on 56 assets across four asset classes (equity indices, commodities, currencies and government bonds) we build time-series momentum strategies using information from our dynamic classifier model and compare with the standard naive time-series momentum where the investor just buy or sell each asset based on specific previous returns. We show that the dynamic classifier approach not only produces better out-of-sample accuracy about the future return directions, but also improves significantly portfolio performances. The model specification using time-varying parameters and DMS to predict future trend signals generated a 52\% increase in  annualized out-of-sample Sharpe ratios compared to the naive approach benchmark. The constant parameter counterpart of the DMS approach also performed quite well, delivering a 44\% Sharpe ratio increase compared to the benchmark. We also show that by the use of DMS or DMA, our dynamic binary classifier was able to explore sudden turning points during the 2009 momentum crash. While the naive time series momentum strategy produced strong losses during the crash period (25\% of cumulative return losses in 16 months), applying dynamic momentum speed selection with time-varying parameters earned 22\% of total cumulative return gains in the same period. Finally, in the same spirit of \cite*{fleming2001economic}, we show that a mean-variance investor will be willing to pay 425 basis points as  annualized management fee to switch from the standard naive time-series momentum strategy to our dynamic classifier approach with time-varying parameters and look-back period selection.

The rest of the paper is organized as follows. Section \ref{naive-momentum} explains the traditional time-series momentum strategy and how to create portfolios based on specific look-back periods. Section \ref{econometric} describes the econometric methodology  behind the dynamic classifier. In Section \ref{Portfolio Allocation} we describe the data used and explore the empirical results of dynamic portfolios strategies, both in terms of out-of-sample predictability and economic performance. In Section \ref{crash_2009} we discuss the economic performance during the well known 2009 crash period and the subsequent years. Finally, Section \ref{Conclusion} concludes.

\section{Standard Time-Series Momentum Strategy}
\label{naive-momentum} 

In this Section we descrive how the most common time-series momentum is performed. The definitions are based on the main literature on time series momentum cited above, in special the work of \cite{moskowitz2012time}. Let $r_{it}$ represent the log-return of security $i$ at month $t$. We can define $Mom_{it}^{L}$ as the momentum measure at time $t$ for security $i$ and look-back period $L$ as:

\begin{equation}
Mom_{it}^{L} = \sum_{l=1}^{L} r_{i, t-l}
\end{equation}

\noindent which is basically the cumulative return
from the previous $L$ periods until time $t-1$. Using the momentum measure, we can find trading signals. For example, if  $Mom_{it}^{L} \geq 0$, it represents a long position (+1) and $Mom_{it}^{L} < 0$ indicates a short position (-1) on asset $i$. It is common practice in the literature to size each asset position so that it has an ex ante annualized volatility target of $\sigma_{tg} = 40\%$ (\citealp{moskowitz2012time}). Hence, the position size is chosen to be $40\%/\sigma_{t}$, where $\sigma_{t}$ is the ex-ante asset volatility estimate. In this manner, the time-series momentum return for asset $i$ at time $t$ will be:

\begin{equation}
r_{it}^{TSMOM,L} =  sign(Mom_{it}^{L} )  \frac{\sigma_{tg}}{\sigma_{it}} r_{it}
\end{equation}

The usual volatility model applied in the literature is the EWMA volatility measure. The annualized volatility can be represented as 

\begin{equation}
\sigma_{it}^{2}= D \sum_{l=0}^{\infty}(1-\delta) \delta^{l}\left(r_{i,t-1-l}-\bar{r}_{i}\right)^{2}
\end{equation}

\noindent where D is the number of observations within a year and $\delta$ is a decay factor\footnote{We consider in this work $\delta=0.97$. We also tested for different values and main results remained similar. Since the focus of this study is to improve return signal predictions, we argue that as soon as the same decay factor is applied to all different strategies, the results are not driven by the volatility measure.} We recognize the simplicity of this volatility measure to capture the right movements of returns volatilities. However, it is important to highlight here that our goal in this study is not to perform volatility timing strategies, but to show performance improvements via signal predictions based on momentums. In order to approximate our study as closely as possible to the format used by the literature and to fairly compare our results we use the same volatility model for all different strategies in the paper. Differently from the work of \cite*{kim2016time}, where the authors argue that time-series momentum strategies are driven by volatility scaling, we show in our results significant portfolio improvements by just modeling return signals instead of volatilities.

Considering a holding period of one month, the return of the overall portfolio diversifying across the $N_{t}$ assets available at time $t$ is simply 

\begin{equation}
r_{pt}^{TSMOM,L} = \frac{1}{N_{t}} \sum_{i=1}^{N_{t}} sign(Mom_{it}^{L} )  \frac{\sigma_{tg}}{\sigma_{it}} r_{it}
\end{equation}

Note that in the standard time-series momentum strategy, return signals are based just on the observation of past returns, i.e., there is an  absence of an econometric model to infer about the correct future return directions and we argue here that it can dramatically reduce the final portfolio performance. Since the standard time-series momentum strategy does not take into account the relationship between past momentum measure and future returns and how it changes over time, the investor is giving up the opportunity to learn about new economic environments to improve signal forecasts and possibly is incurring in misleading trading positions. This is why, for now on, we will refer to the standard time-series momentum strategy as naive.

\section{Dynamic classifier method}
\label{econometric} 

Now we describe in details the dynamic classifier model used in this paper. It is mainly inspired by the work of \cite{mccormick2012dynamic} and can be viewed as an econometric substitute for the traditional naive time-series momentum approach. As before, the investor wishes to antecipate the future direction of asset returns in order to set long or short positions. Therefore, we still have a binary classification problem. The great difference now is that decisions will be based on a statistical model that is able to better digest today's information complexity to infer about the future return direction and improve trading decisions.

The econometric method is based on a dynamic logistic regression model for each individual return series. The model is written in state-space form, where $s_{t}$ represent a binary response of a individual asset return, i.e., $s_{t}=1$ when returns are greater or equal to zero and $s_{t}=0$, otherwise. Let $\boldsymbol{x_{t}}$ be a $d$-vector containing a set of possible momentum measures as predictors for the specific asset return signal. Then: 

\begin{equation}
s_{t} \sim Bernoulli(p_{t})
\end{equation}

\noindent where 

\begin{equation}
logit(p_{t}) = log\left(\frac{p_{t}}{1-p_{t}}\right) = \boldsymbol{x}_{t}^{\prime}\boldsymbol{\theta}_{t}
\end{equation}

\begin{equation}
\boldsymbol{\theta}_{t} = \boldsymbol{\theta}_{t-1} + \boldsymbol{\omega}_{t}, \quad  \boldsymbol{\omega}_{t} \sim N\left(0, \boldsymbol{W}_{t} \right) 
\end{equation}

\noindent where $p_{t}$ is the probability of a positive return signal and the $d$-vector $\boldsymbol{\theta}_{t}$ contains regression coefficients representing the relationships among momentum and future return direction and also an intercept coefficient. Note that coefficients are allowed to evolve over time as random-walks. Here, we initially consider a single arbitrary model containing a specific set of momentum predictors  $\boldsymbol{x_{t}}$. Later, in the next section we explore our dynamic momentum learning procedure where different models with different momentum predictors are considered. For now, consider the existence of $K$ possible look-back periods we may use and let $\boldsymbol{x_{t}}$ contain only one of the $2^{K}$ possible combination of predictors we may include in the particular model.\footnote{Hence, for instance, we could have $\boldsymbol{x_{t}} = \left[1,   Mom_{it}^{L=1}, Mom_{it}^{L=6}, Mom_{it}^{L=12}\right]$ or even  $\boldsymbol{x_{t}} = \left[1,   Mom_{it}^{L=2}, Mom_{it}^{L=8}\right]$.}

Let $\mathcal{D}_{t-1}$ represents the whole information available until time $t-1$, i.e., $\mathcal{D}_{t-1} = s_{1},\dots, s_{t-1}$. Hence, the posterior distribution for coefficients at time $t-1$ is

$$
\boldsymbol{\theta}_{t-1} \mid \mathcal{D}_{t-1} \sim N\left(\boldsymbol{m}_{t-1}, \boldsymbol{C}_{t-1}\right)
$$

\noindent where $\boldsymbol{m}_{t-1}$ and $\boldsymbol{C}_{t-1}$ represent the posterior mean and covariance for $\boldsymbol{\theta}_{t-1}$ at time $t-1$ . The prediction equation for time $t$ given information until time $t-1$ is given by

$$
\boldsymbol{\theta}_{t} \mid  \mathcal{D}_{t-1} \sim N\left(\boldsymbol{a}_{t}, \boldsymbol{R}_{t}\right)
$$

\noindent where $\boldsymbol{a}_{t} = \boldsymbol{m}_{t-1}$ and by the use of a discount factor $0 < \lambda_{t} \leq 1$ we can obtain the predicted covariance matrix of states as $\boldsymbol{R}_{t}= \frac{\boldsymbol{C}_{t-1}}{\lambda_{t}}$. The use of discounting methods simplifies estimation and is widely applied in the bayesian literature. It can be viewed as a way to discount more heavily past information. A discount factor lower than 1 imposes time-variation in coefficients while $\lambda_{t}= 1$ set coeficients to be constant (see \citealp{west2006bayesian}, \citealp{prado2010time} and \citealp*{raftery2010online} for details about discounting methods).

After computing the prior state distribution for time $t$, we are able to generate a return signal prediction for the specific asset as:

\begin{equation}
\widehat{s}_{t|t-1} = \frac{1}{1+e^{-\boldsymbol{x}_{t}^{\prime}\boldsymbol{a}_{t}}}
\label{forecast}
\end{equation}

\noindent which will be used to portfolio decisions as we explain in Section \ref{Portfolio Allocation}. 

At time $t$, the investor observes $s_{t}$ and is able to update her state estimates by the Bayes' rule:

\begin{equation}
p\left(\boldsymbol{\theta}_{t} \mid \mathcal{D}_{t}\right) \propto p\left(s_{t} \mid \boldsymbol{\theta}_{t}\right) p\left(\boldsymbol{\theta}_{t} \mid \mathcal{D}_{t-1}\right)
\label{posterior}
\end{equation}

\noindent which is simply the product of the likelihood at time $t$ and the prediction equation for $\theta_{t}$ defined above. Since Equation (\ref{posterior}) is not available in closed-form, \cite{mccormick2012dynamic} approximate this posterior with a normal distribution. Let $l(\boldsymbol{\theta}_{t}) = log p\left(s_{t} \mid \boldsymbol{\theta}_{t}\right) p\left(\boldsymbol{\theta}_{t} \mid \mathcal{D}_{t-1}\right)$ and $Dl(\boldsymbol{\theta}_{t})$ and $D^{2}l(\boldsymbol{\theta}_{t})$ being its first and second derivative. The mean of the approximate distribution will be the mode of Equation (\ref{posterior}) and its estimate will be given by

\begin{equation}
\boldsymbol{m}_{t} = \boldsymbol{a}_{t} - D^{2}l(\boldsymbol{\theta}_{t})^{-1} Dl(\boldsymbol{\theta}_{t})
\label{mode} 
\end{equation}

\noindent and the state variance is updated by $\boldsymbol{C}_{t} = - D^{2}l(\boldsymbol{\theta}_{t})^{-1}$.  

In order to apply DMA or DMS (see below) and tune discount factors, the predictive likelihood will be taken into account:

\begin{equation}
p\left(s_{t} \mid \mathcal{D}_{t-1}\right)=\int_{\boldsymbol{\theta}_{t}} p\left(s_{t} \mid \boldsymbol{\theta}_{t}, \mathcal{D}_{t-1}\right) p\left(\boldsymbol{\theta}_{t} \mid \mathcal{D}_{t-1}\right) d \boldsymbol{\theta}_{t}
\end{equation}

However, since this integral is not available in closed form, a Laplace approximation is used such that:

\begin{equation}
\begin{array}{l}
p\left(s_{t} \mid \mathcal{D}_{t-1}\right) \approx(2 \pi)^{d / 2}\left|\left\{D^{2}l\left(\boldsymbol{\theta}_{t}\right)\right\}^{-1}\right|^{1 / 2} p\left(s_{t} \mid \mathcal{D}_{t-1}, \boldsymbol{\theta}_{t}\right)  p\left(\boldsymbol{\theta}_{t} \mid \mathcal{D}_{t-1}\right)
\label{laplace}
\end{array}
\end{equation}

\noindent which makes computation much faster, since no expensive simulation schemes are required. In order to tune $\lambda_{t}$, we propose a grid of values for $\lambda_{t}$ and sequentially select over time the one such that Equation (\ref{laplace}) is maximized. In our empirical section below, we use $\lambda_{t}$ $\in$ $\{0.98, 0.99, 1 \}$ for the time-varying parameter (TVP) setting. Hence, our approach is able to induce higher or lower degree of variability in coefficients if it is empirically suited. For the constant parameter (CP) case, no discounting is applied, so we fix $\lambda=1$ for all periods of time.

\subsection{Dynamic momentum learning}

Considering the uncertainty around which look-back period brings more information about future direction of returns, now we explain how momentum (speed) uncertainty can be inserted in our dynamic classifier model. Suppose there are $K$ possible look-back periods an investor may consider for a time-series momentum strategy for a specific asset. Since there is uncertainty about the amount of economic information each look-back period may provide to infer the future direction of returns and what is the best speed (or combination of speeds), the investor is faced with the problem of momentum uncertainty. How is an investor able to understand the complexity of the trend structure just by looking at past returns? There are several different paths that may define a positive or negative trend. For instance, one asset may have a slower (long period) positive momentum trend, but a fast (short period) negative momentum. Also, there are cases of long and short positive (negative) momentum trends, but negative (positive) intermediate trend. We argue here that those patterns are changing over time, since the financial market is continuously adapting to new environments. Therefore, the idea of dynamic momentum learning is to compute all the $M = 2^{K}$ possible models\footnote{Hence, the model space will be determined by different momentum measures (look-back periods). In our work we exclude the case of zero predictors, so actually we consider $M = 2^{K} - 1$ possible models.} in parallel and assign dynamic model probabilities for each one in such way that the investor is able to sequentially learn from past model mistakes, switching to model settings that are performing better in the recent past or combining all of them weighting by their model probabilities.

Denote $\pi_{t-1 | t-1, i} = p(\mathcal{M}_{i} \mid  \mathcal{D}_{t-1})$ as the posterior probability of a model $i$ with a specific subset of momentum predictors at time $t-1$.\footnote{Initially, we assign equal probabilities for all $M=2^{K}$ different models: $\pi_{0 | 0, i} = \frac{1}{M}$ for all models $i$.} Following \cite{raftery2010online} and \cite{mccormick2012dynamic}, the predicted probability of the model $i$ given all the data available until time $t-1$ can be expressed as:

 \begin{align}
 \pi_{t | t-1, i}=\frac{\pi_{t-1 | t-1, i}^{\alpha}}{\sum_{l=1}^{M} \pi_{t-1 | t-1, l}^{\alpha}},
\end{align}

\noindent where $0 \leq \alpha \leq 1$ is another discounting (forgetting) factor. The main advantage of using $\alpha$ is avoiding the computational burden associated with expensive MCMC schemes to simulate the transition matrix between possible models over time. This approach has also been extensively used in the Bayesian econometric literaure in the last decade (\citealp{koop2013large}, \citealp*{zhao2016dynamic}, \citealp*{lavine2020adaptive} and \citealp*{beckmann2020exchange}). After observing new data at time $t$, we update model probabilities following a simple Bayes’ rule:

\begin{align}
 \pi_{t | t, i}=\frac{\pi_{t | t-1, i}p_{i}\left(s_{t} \mid  \mathcal{D}_{t-1}\right)}{\sum_{l=1}^{M} \pi_{t | t-1, l}p_{l}\left(s_{t} \mid \mathcal{D}_{t-1}\right)}.
\end{align}

\noindent which is the posterior probability of model $i$ at time $t$ and $p_{i}(s_{t} \mid \mathcal{D}_{t-1})$ is the predictive density of model $i$ evaluated at $s_{t}$. Note that the predictive densities have already been computed as we have shown in Equation (\ref{laplace}), which implies that no extra computations are required here to update model probabilities. Hence, upon the arrival of a new data point, the investor is able to measure the performance of each model $i$ and to assign higher probabilities for those models that generate better out-of-sample performance. 

One possible interpretation for the forgetting factor $\alpha$ is through its role to discount past performance. Combining the predicted and posterior probabilities, we can show that

\begin{equation}
\pi_{t \mid t-1, i} \propto \prod_{l=1}^{t-1}\left[p_{i}\left(s_{t-l} \mid  \mathcal{D}_{t-l-1}\right)\right]^{\alpha^{l}}.
\label{eq:pis}
\end{equation}

Since $0 < \alpha \leq 1$, Equation (\ref{eq:pis}) can be viewed as a discounted predictive likelihood, where past performances are discounted more than recent ones. It implies that models that generated higher out-of-sample performance in the recent past will receive higher predictive model probabilities. The recent past is controlled by $\alpha$, since a lower $\alpha$ discounts more heavily past data and generates a faster switching behavior between models over time and $\alpha = 1$ represents no forgetting information. The value of $\alpha_{t}$ is sequentially selected over time such that it maximizes the average predictive likelihood over all different model:

\begin{equation}
\alpha_{t}  = \arg \max _{\alpha_{t}}  \sum_{l=1}^{M} \pi_{t | t-1, l}p_{l}\left(s_{t} \mid \mathcal{D}_{t-1}\right)
\end{equation}

Similar to the discount factor $\lambda_{t}$, we propose a grid of values $\alpha_{t} \in \{0.99,1\}$ such that the model can switch between forgetting and no-forgetting information over time. 

Using predictions for each individual model $i$, $\widehat{s}_{t|t-1}^{i}$, we compute the dynamic model average prediction ($\widehat{s}_{t|t-1}^{DMA}$) weighting by each individual model probability:

\begin{equation}
\widehat{s}_{t|t-1}^{DMA}  = \sum_{l=1}^{M} \pi_{t | t-1, l}  \widehat{s}_{t|t-1}^{l}
\label{eq_dma}
\end{equation}

\noindent while dynamic model selection (DMS) is applied by simply selecting the model $i$ with the highest predicted model probability, $\pi_{t | t-1, i}$, for period $t$, . 

For each asset available, the investor can apply the whole procedure described above and classify long or short position for individual assets based on different  out-of-sample signal predictions. In the next section we explain in details the time-series momentum strategy using the dynamic classifier approach and describe the data used in our empirical study.

\section{Dynamic Portfolio Allocation}
\label{Portfolio Allocation}

Now we discuss how to incorporate output informations from the dynamic classifier approach as inputs in a dynamic time-series momentum strategy. Supposing there are $N_{t}$ assets available for investing at time $t$, instead of considering $sign(Mom_{it}^{L} )$ as a signal classifying long or short positions, the investor will use the DMA or DMS classification prediction for each individual asset to generate trading signals for her portfolio. Hence, if $\widehat{s}_{t|t-1} \geq c$ it indicates a long position ($sign(\widehat{r}_{t|t-1}) = +1$) and if $\widehat{s}_{t|t-1} < c$ we have a short position ($sign(\widehat{r}_{t|t-1}) = -1$) in that specific asset, where $c$ represents a cutoff selected by the investor. The most direct and simple cutoff choice is to consider $c = 0.5$ or to apply a sequential grid search to maximize out-of-sample accuracy for each individual asset. In our empirical results we show results using both approaches. Therefore, the return of the overall time-series momentum portfolio using DMA will be given by

\begin{equation}
r_{pt}^{TSMOM - DMA} = \frac{1}{N_{t}} \sum_{i=1}^{N_{t}} sign(\widehat{s}_{t|t-1,i}^{DMA}) \frac{\sigma_{tg}}{\sigma_{it}} r_{it}
\label{dma_port}
\end{equation}

\noindent and the return of the time-series momentum portfolio with signals obtained from the DMS procedure will be given by

\begin{equation}
r_{pt}^{TSMOM - DMS} = \frac{1}{N_{t}} \sum_{i=1}^{N_{t}} sign(\widehat{s}_{t|t-1,i}^{DMS}) \frac{\sigma_{tg}}{\sigma_{it}} r_{it}
\label{dms_port}
\end{equation}

\noindent where results with constant and time-varying parameters are shown in the empirical section. We also show portfolio performances when, instead of using model combination or selection, we use models with single predictors. For example, when we display results for TVP-12m, we are referring to a model using the twelve month momentum measure as the only predictor and time-varying parameters are allowed. In this case, portfolios are formed as in Equations (\ref{dma_port}) and (\ref{dms_port}), where signals are coming from this specific single predictor model. 

Using these simple diversified portfolios, we  follow the majority of the works related to time-series momentum strategies. It allows us to evaluate the real economic improvements due exclusively to  our dynamic classifier method using momentum combination or momentum selection compared to the standard time-series momentum classification and not due to volatility timing effects.\footnote{For correlation/volatility timing and the benefits of risk parity allocations in time-series momentum strategies we refer to \cite{baltas2020demystifying} and \cite{rubesam2020long}. \cite{kim2016time} also explore the effects of volatility scaling on time-series momentum strategies. } We let volatility timing in time-series momentum strategies as a future research extension to our approach.

Therefore, for each period of time, using the method described above, the investor uses signal forecasts for each individual asset $\widehat{s_{it}}$ as inputs in the dynamic portfolio, sizing each position based on ex ante volatilities as in Equation (\ref{dma_port}). Since the number of assets available for investors at the beginning of the sample does not comprise the entire data sample, it is important to notice that the number of assets which enters the portfolio $N_{t}$ varies over time.  

In our empirical study we will consider several specification in order to compare the benefits of using time-varying parameters, model averaging and model selection to the naive time series momentum (Naive-TSMOM). We show statistical and economic results for Naive-TSMOM strategies considering fixed look-back periods of 1, 2, 4, 6, 8, 10 and 12 months. Then, we also report performances when using this same fixed look-back periods when the trading signals are obtained from our dynamic classifier model. That is, when there is no model uncertainty and a single momentum measure is applied within our econometric model. In this cases, we divide results for look-back periods of  1, 2, 4, 6, 8, 10 and 12 months for constant parameters (CP: $\lambda = 1$) and also when time-varying parameters are allowed (TVP: $\lambda_{t} \in \{0.98, 0.99,1\}$). The CP specification can be viewed as a recursive static logistic regression. Finally, we show results for both CP and TVP when DMA and DMS approaches are applied, meaning that the investor is dynamically learning momentum speed probabilities and averaging or selecting predictions based on those probabilities.

\subsection{Data}

Following \cite{rubesam2020long}, we use data of continuous prices for 56 futures contracts downloaded from Refinitiv/Datastream for the
period from January 1980 to September 2020. The data covers 12 developed market equity index futures, 25 commodity futures, 11 developed sovereign bond futures and 8 currency pairs futures. The contracts are rolled over the last trading day of the expiry month and adjusted at the roll date to avoid artificial returns. Since our methodology is inherently built for one-step ahead predictions and daily portfolio rebalancing prevents the best exploitation of longer momentum effects and increases transaction costs, we decide to follow the great majority of the literature and use monthly log-returns in our study. We consider one month as holding period, that is, we rebalace portfolios in a monthly basis. Tables (\ref{summary}) and (\ref{summary_2}) in the Appendix provide a statistical summary of future contracts and its start dates.

In our analysis we use the first three years (36 months) of data for each individual asset as a training period for our models and use the rest of the subsequent periods as an out-of-sample evaluation period. Hence, as soon as a new asset is available in the data set, it enters in the portfolio just three years later. Although the naive time series momentum strategy does not rely on a econometric approach and does not require a training period, in our empirical results we also discard the same initial sample window in order to fairly compare all approaches with the same data set.

\subsection{Out-of-Sample Predictability}
\label{predictability}

We analyse how the proposed dynamic econometric specifications perform in terms of out-of-sample forecasting by comparing models via mean absolute errors (MAE). In order to provide an overall metric considering all different asset returns and its particular different periods within the real portfolio application, we decided to stack all asset returns as they were a single series and average absolute error considering the sum of the test sample periods of each asset. More specifically, consider that for each asset $i$ the training period ends at month $T_{train, i}$ and its number of months in the test sample is $T_{test, i}$, hence we compute MAE of an econometric approach $j$ as:

\begin{equation}
MAE^{j} = \frac{\sum_{i} \sum_{t=(T_{train, i}+1)}^{T} \left| \widehat{s_{it}} - s_{it}\right|}{\sum_{i} T_{test, i}} 
\label{BS}
\end{equation}

\noindent where $T$ is the last month of our sample (September, 2020). Differently from an econometric forecasting model, the naive strategy does not provide a probability prediction per se, so the investor should buy or sell based only on the sign of past accumulated returns. Therefore, we compare absolute errors within different econometric models, where the chosen benchmark is the binary classifier using the twelve months momentum as the only predictor and using constant parameters (CP-12m). We report relative percentage performance improvements

\begin{equation}
(\%) MAE^{j} = 1 - \frac{MAE^{j}}{MAE^{CP-12m}} 
\label{BS}
\end{equation}

Hence, positive numbers represent a percentage reduction in terms of out-of-sample forecasting error compared to the statistical model using just the look-back period of 12 months and constant parameters. 

Since the forecasted value from our dynamic classifier approach is an estimate for the probability of a positive return, a true positive (TP) occurs when the specific model forecasted a probability of positive returns greater or equal than a cutoff $c$ and it coincides with a positive realized return, while a true negative (TN) occurs when the model forecasted a value lower than the cutoff $c$ and it coincides with a negative realized return. At the other hand, false positives (FP) and false negative (FN) represent the case where the realized return is the opposite to what the forecasted values were indicating. As in \cite{jiang2020re}, we compute classification accuracy as:

$$
Accuracy = \frac{TP + TN}{TP + TN + FP + FN}
$$

\noindent where again we consider the sum of TP, TN, FP and FN for all assets available to compute total accuracy for each strategy approach. In our main results we consider a cutoff $c=50\%$. However, as a robustness test, we also report results considering cross-validation procedures to sequentially select the best cutoff among different values in two grids and for each asset individually over time. The first cross-validation procedure ($CV_{1}$) considers a grid $c \in \{0.49, 0.491, ..., 0.509, 0.51\}$ and the second cross-validation procedure ($CV_{2}$) considers $c \in \{0.45, 0.46, ..., 0.54, 0.55\}$. Those are reasonable grids of values, since much higher or lower values tend to induce mislieading trading signals. The idea of the grid search CV is to select the $c$ such that it produces the highest accuracy rate over time. The CV is repeated once a month and we use the last three years of out-of-sample accuracy observations to select the best $c$. In this sense, such smaller grid of values also avoids aditional computational burden, since the procedure is repated several times for each asset available. 
Inspired by a Bayesian Decision Theory perspective, we show in the Appendix an additional portfolio exercise where the investor maximizes an expected utility where probabilities of different scenarios are coming from our classifier models. Table (\ref{stats}) below shows the results for forecast performance without CV ($c=50\%$), for both CV procedures and the MAE metric.

\begin{table}[!htbp] \centering 
  \caption{Out-of-Sample Forecast performance (\%)} 
  \label{stats} 
\begin{tabular}{@{\extracolsep{5pt}}l ccccc} 
\\[-1.8ex]\hline 
\hline \\[-1.8ex] 
 & Acc ($c=50\%$) & Acc ($CV_{1}$) & Acc ($CV_{2}$) & MAE \\ 
\hline \\[-1.8ex] 
  &  &  &  \\ 
  &  & \textbf{CP} &  \\ 
  \hline \\[-1.8ex] 
DMA & $53.1$ & $53.0$ & $52.9$ & $0.342$ \\ 
DMS & $53.3$ & $53.2$ & $53$ & $0.470$ \\ 
1m & $52.7$ & $52.7$ & $52.8$ & $0.067$ \\ 
2m & $52.5$ & $52.6$ & $52.8$ & $-0.017$ \\ 
4m & $52.7$ & $52.6$ & $52.5$ & $0.097$ \\ 
6m & $52.7$ & $52.7$ & $52.5$ & $0.038$ \\ 
8m & $53.0$ & $52.6$ & $52.6$ & $0.162$ \\ 
10m & $52.9$ & $52.8$ & $52.3$ & $0.083$ \\ 
12m & $52.4$ & $52.2$ & $52.3$ & $0$ \\ 
\hline \\[-1.8ex] 
 &  &  &  \\ 
  &  & \textbf{TVP} &  \\ 
  \hline \\[-1.8ex] 
DMA & $53.1$ & $53.1$ & $53$ & $0.407$ \\ 
DMS & $53.1$ & $53.2$ & $53.1$ & $0.502$ \\ 
1m & $52.8$ & $52.8$ & $52.9$ & $0.176$ \\ 
2m & $52.4$ & $52.7$ & $53$ & $-0.6$ \\ 
4m & $52.8$ & $52.6$ & $52.8$ & $0.098$ \\ 
6m & $52.7$ & $52.6$ & $52.9$ & $0.045$ \\ 
8m & $52.5$ & $52.8$ & $52.7$ & $0.138$ \\ 
10m & $52.5$ & $52.6$ & $52.3$ & $0.056$ \\ 
12m & $52.2$ & $52.4$ & $52.2$ & $0.012$ \\ 
\hline \\[-1.8ex] 
 &  &  &  \\ 
  &  & \textbf{Naive-TSMOM} &  \\
  \hline \\[-1.8ex] 
1m & $50.1$ &  &  &  \\ 
2m & $51.0$ &  &  &  \\ 
4m & $51.2$ &  &  &  \\ 
6m & $51.2$ &  &  &  \\ 
8m & $51.4$ &  &  &  \\ 
10m & $52.2$ &  &  &  \\ 
12m & $52.3$ &  &  &  \\ 
\hline \\[-1.8ex] 
\end{tabular} 
\begin{tablenotes}
      \small 
 \item       \textit{The table reports out-of-sample forecast performance from classifier models for constant parameters (CP) and time-varying parameters (TVP), considering single predictors (1m, 2m, 4m, ..., 12m) or applying dynamic model averaging (DMA) or dynamic model selection (DMS) with all predictors in the model space. $CV$ stems for cross-validation, where the grid from $CV_{1}$ range from 0.49 to 0.51 and $CV_{2}$ ranges from 0.45 to 0.55. The bottom panel shows accuracy from the naive time-series momentum strategy (Naive-TSMOM). }
    \end{tablenotes}
\end{table}

From Table (\ref{stats}), the column referring to MAE shows forecast error reductions for different models compared to the CP classifier approach using a twelve month momentum predictor. Both DMA and DMS demonstrate higher improvements, specially when TVP are allowed. We can note that DMS has a slightly better performance compared to DMA and none of the models using a single preditor was able to outperform model selection or combination. Focusing on classification accuracy, the first column of Table (\ref{stats}) shows results when a single cutoff $c=50\%$ is applied for all assets and for all periods of time. Since this accuracy can be obtained from a signal, the Naive-TSMOM is able to infer about this trading signals by just looking to past momentum measures. The bottom panel of the table shows a good performance for Naive-TSMOM when longer look-back periods are considered. Indeed, the Naive-TSMOM of twelve months generated a forecast accuracy of 52.3\%. This is in line with the empirical academic research, where the look-back period of twelve months is the main setting for many studies (\citealp{moskowitz2012time} and \citealp{rubesam2020long}).

The first and second panels of Table (\ref{stats}) show that by the use of a binary classifier approach, forecast accuracy can be considerably improved. It is interesting to note that i) for model settings with a single momentum predictor and short look-back periods, the performance tend to be better than longer look-back periods, the opposite to what is observed from Naive-TSMOM strategies, ii) single predictor models with constant parameters and longer look-back periods tend to perform better than their time-varying parameter counterparts, while for shorter look-back periods time-varying parameters are slightly better than their constant parameter counterpart, and iii) model combination and selection were able to significantly increase out-of-sample classification accuracy. For both CP and TVP, DMA and DMS delivered accuracies higher than 53\%, with DMS-CP being of 53.3\%. In the next sections we show that, although DMS-CP performed slightly better than DMS-TVP in terms of total accuracy, the TVP version has greater versatility to recognize sudden turining points, reducing drawdowns and performing better during bad periods for time-series momentum strategies, which ends up improving its final economic performance.

Although results in Table (\ref{stats}) may still seem low, differently from other binary classification applications and literatures, it is important to remember that in the context of return predictability where forecasting is a huge challange, any tiny accuracy improvement can be translated in strong economic improvements across time for the investor (\citealp{chinco2019sparse}, \citealp{gu2020empirical} and \citealp{jiang2020re}). In fact, \citealp{jiang2020re} show evidences that small accuracy gains in the order of 1\% is able to be translated into considerable Sharpe ratio gains for trading strategies based on these predictions.

In order to test the robustness of results for different cutoffs, we let them change over time and for each individual asset. The second and third columns from Table (\ref{stats}) show results when $c$ are coming from a cross-validation procedure, where each column utilizes a different grid of possible values, as explained above. In fact, the results are still robust with small improvements for the TVP in the $CV_{1}$ case for longer look-back periods and the DMS approach, whereas for the CP counterpart the results are slightly improved for shorter look-back periods and harmed for longers and model combination and selection. In the $CV_{2}$, there is a small improvement for short momentums in both the TVP and CP models. In general, the results are still quite similar when $c=50\%$ and model combination and selection continue to show accuracy gains compared not just to single predictor models but specially to the Naive-TSMOM strategy.

\subsection{Economic Performance}

At the end of each month, the investor observes the data available and predicts the future directions of returns for the end of the next month. After obtaining forecasting outputs (and calibrating signals cutoffs when applying the cross-validation procedure), the investor is then able to use the prediction outputs as inputs in a portfolio allocation problem, sequentially rebalancing her portfolio. Using predictions from each model setting, we build portfolios as in Equations (\ref{dma_port}) and (\ref{dms_port}). 

In order to show economic improvements for investors, we show important measures of portfolio performance such as annualied mean excess returns (Mean), volatilities (Vol.), Maximum Drawdowns (Max DD) and Sharpe Ratios (SR). The latter is commonly used among practicioniers in the financial market and by academics. Despite its popularity, SR is an unconditional measure and is not well suited for dynamic allocations with time-varying and sequential predictions (see \citealp{marquering2004economic}). Also, they do not take into account the investor risk aversion. In order to overcome this problems and improve our model comparisons, we follow \cite{fleming2001economic} and provide a measure of economic utility for investors. We compute ex-post average utility for a mean-variance investor with a quadratic utility and calculate the performance fee that an investor will be willing to pay to switch from the standard time-series momentum strategy to the dynamic classifier method (DCM):

\begin{equation}
  \small \small \sum_{t=0}^{T-1}\{(R_{p, t+1}^{DCM}-\Phi)-\frac{\gamma}{2(1+\gamma)}(R_{p, t+1}^{DCM}-\Phi)^{2}\} =  \sum_{t=0}^{T-1}\{R_{p, t+1}^{Naive-12m}-\frac{\gamma}{2(1+\gamma)}(R_{p, t+1}^{Naive-12m})^{2} \}
  \label{eq_utility}
\end{equation}

\noindent where $\gamma$ is the investor's degree of relative risk aversion, $R_{p,t}^{DCM}$ is the gross return of the specific DCM portfolio and $R_{p,t}^{Naive-12m}$ is the gross return from the Naive-TSMOM strategy portfolio when a look-back period of twelve months is considered. As in \cite{fleming2001economic}, we report our estimates of $\Phi$ as annualized management fees in basis points using $\gamma = 10$ as risk aversion. Notice that $\Phi$ is computed by equating the average utility from the investor applying the Naive-12m strategy with the average utility of the DCM portfolio (or any other alternative specification).

In an effort to bring our results as closer as possible to a real world example, we follow \cite{baltas2020demystifying} and \cite{rubesam2020long} and report results net of transaction costs. Both rebalancing and rollover costs are taken into account. Each asset class will rely on different transaction costs following the same values in basis points reported in \cite{baltas2020demystifying}.

\begin{table}[!h] \centering 
  \caption{Economic Performance of TSMOM strategies ($c = 50\%$)} 
  \label{economic_1} 
\begin{tabular}{@{\extracolsep{5pt}}l ccccccc} 
\\[-1.8ex]\hline 
\hline \\[-1.8ex] 
 & Turnover & Mean & Vol. & Max.DD & SR & $\Phi$ \\ 
\hline \\[-1.8ex] 

&  &  &  &  &  & \\ 
&  &  & \textbf{CP} &  &  & \\ 
  \hline \\[-1.8ex] 
DMA & $67.9$ & $10.5$ & $10.0$ & $15.1$ & $1.05$ & $240.4$ \\ 
DMS & $79$ & $11.7$ & $10.0$ & $17$ & $1.17$ & $366.3$ \\ 
1m & $64$ & $8.1$ & $10.0$ & $22.1$ & $0.81$ & $-7.9$ \\ 
2m & $53.2$ & $6.2$ & $10.0$ & $24.2$ & $0.62$ & $-187.1$ \\ 
4m & $50.2$ & $8.0$ & $10.0$ & $20.1$ & $0.80$ & $-17.9$ \\ 
6m & $46$ & $8.3$ & $10.0$ & $18$ & $0.83$ & $12.6$ \\ 
8m & $44.4$ & $9.3$ & $10.0$ & $14.4$ & $0.93$ & $121.9$ \\ 
10m & $42.3$ & $9.3$ & $10.0$ & $16.9$ & $0.93$ & $122.5$ \\ 
12m & $43.1$ & $7.0$ & $10.0$ & $15.7$ & $0.70$ & $-110.6$ \\ 
\hline \\[-1.8ex] 
&  &  &  &  &  & \\ 
&  &  & \textbf{TVP} &  &  & \\ 
  \hline \\[-1.8ex] 
DMA & $85.6$ & $10.6$ & $10.0$ & $13.1$ & $1.06$ & $254.5$ \\ 
DMS & $101.3$ & $12.3$ & $10.0$ & $14.8$ & $1.23$ & $425.1$ \\ 

1m & $79.6$ & $8.6$ & $10.0$ & $15.3$ & $0.86$ & $42.7$ \\ 
2m & $67$ & $6.0$ & $10.0$ & $25.9$ & $0.60$ & $-210.6$ \\ 
4m & $58.1$ & $7.5$ & $10.0$ & $26.5$ & $0.75$ & $-62.1$ \\ 
6m & $53.4$ & $9.0$ & $10.0$ & $21.2$ & $0.90$ & $86.0$ \\ 
8m & $55.8$ & $8.4$ & $10.0$ & $16.8$ & $0.84$ & $22.8$ \\ 
10m & $54.4$ & $7.3$ & $10.0$ & $20.2$ & $0.73$ & $-82.8$ \\ 
12m & $55.3$ & $6.1$ & $10.0$ & $20.1$ & $0.61$ & $-203.6$ \\ 
\hline \\[-1.8ex] 
&  &  &  &  &  & \\ 
&  &  & \textbf{Naive} &  &  & \\ 
  \hline \\[-1.8ex] 
1m & $305.9$ & $3.9$ & $10.0$ & $23.0$ & $0.39$ & $-411.5$ \\ 
2m & $207.2$ & $5.0$ & $10.0$ & $12.2$ & $0.50$ & $-307.1$ \\ 
4m & $144.6$ & $4.9$ & $10.0$ & $15.6$ & $0.49$ & $-313.8$ \\ 
6m & $114.4$ & $4.2$ & $10.0$ & $16.7$ & $0.42$ & $-388.0$ \\ 
8m & $98.1$ & $5.2$ & $10.0$ & $19.3$ & $0.52$ & $-288.7$ \\ 
10m & $82.5$ & $7.2$ & $10.0$ & $20.7$ & $0.72$ & $-89.0$ \\ 
12m & $77.1$ & $8.1$ & $10.0$ & $25.3$ & $0.81$ & $0.0$ \\ 
\hline \\[-1.8ex] 
\end{tabular} 
\begin{tablenotes}
      \small 
  \item      \textit{The table reports economic performance from classifier models using constant parameters (CP) and time-varying parameters (TVP), considering single predictors (1m, 2m, 4m, ..., 12m) or applying DMA or DMS with all predictors in the model space. The bottom panel shows performances from naive time-series momentum strategies with different look-back periods. Results consider a classifier cutoff of $c=50\%$ and strategies are scaled to an ex-post annualized volatility of 10\%.}
    \end{tablenotes}
\end{table}

First, we show in Table (\ref{economic_1}) annualized results when no CV is performed to obtain the cutoffs. Hence, $c = 0.5$ is selected for all periods of time and for all assets. All time-series momentum strategies are scaled to an ex-post annualized volatility of 10\%. Focusing in the bottom panel of Table (\ref{economic_1}), which refers to the Naive TSMOM strategy, we note similar performances to other studies. When using $L=12$ months, the strategy performed particularly well over the last decades. It delivered a SR of 0.81 and 8.1\% of annualized average excess return, both metrics being higher than using shorter look-back periods. For instance, a naive strategy using 6 months look-back period would deliver almost the half of the traditional 12-month look-back period strategy. Additionally, shorter look-back period strategies tend to generate much higher portfolio turnovers, which also harm final performance due to transaction costs. Within the Naive-TSMOM group, all shorter look-back period strategies have presented lower maximum drawndowns than the 12-month strategy. Finally, in terms of utility gains, an investor applying any shorter momentum speed would be willing to pay an annualized managment fee from 89 to 411 bps to use the naive-approach with 12-month look-back period.

When we focus on econometric approaches, the results differ to the naive approach. We first notice that using the 12-month momentum as a single predictor does not imply a better portfolio performance and the 1-month predictor was able to deliver strong performances, in a comparable magnitude to the 12-month naive benchmark. Single predictor models tend to induce much lower portfolio turnovers than the naive approach, specially when constant parameters are applied. In fact, the constant parameter group delivered similar or even better results compared to the time-varying parameter case, except for 1 and 6-month momentum predictors. However, when model selection or averaging are applied to different momentum predictors, models with time-varying parameters appear with superior performance compared to its CP counterpart. The DMS-TVP approach was able to dramatically improve portfolio performance compared to the naive approach, generating an annualized Sharpe ratio of 1.23, representing an increase of more than 50\%, while its CP counterpart delivered a SR of 1.17. For both CP and TVP panels, DMA performed better than any individual momentum predictors but worse than model selection. One important aspect of model selection and combination is the ability to reduce maximum drawdowns with higher reductions when TVP are allowed. While the naive benchmark suffered a drastic maximum drawdown of 25.3\% (see Section \ref{crash_2009} for a deeper discussion on drawdowns and the 2009 momentum crash), the DMA-TVP suffered a maximum drawdown of 13.1\%, almost half the size of the losses obtained from the naive benchmark. In terms of monthly turnover, both model selection and combination increase position changes over time compared to single predictor models and the naive approach, specially for TVP models. However, this turnover increase is more than compensated with higher accuracy and portfolio returns.

 Our results confirm that dynamically combining economic informations from different look-back periods using DMA or sequentially selecting the best look-back periods by DMS improves portfolio performance, being consistent with our previous out-of-sample accuracy results. The last column of Table (\ref{economic_1}) shows that a mean-variance investor will be willing to pay 425 bps as annualized management fee to switch from the naive 12-month time-series momentum strategy to our dynamic classificer approach with momentum speeds selection and time-varying coefficients. The economic performances are still strong when using DMA and/or constant parameters.

\begin{figure}[!h]
\begin{center}
\caption{Sharpe Ratios by asset classes}
\label{sr_classes}
\includegraphics[width=0.8\textwidth]{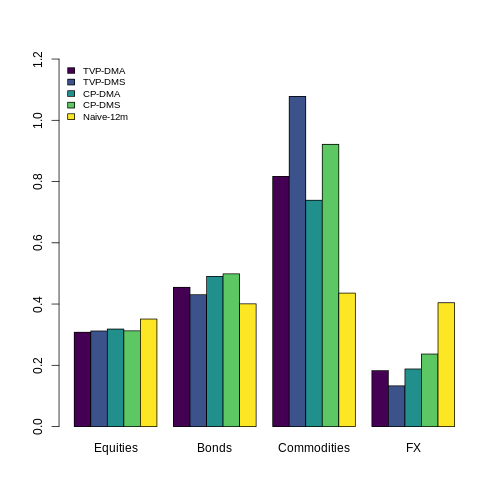}
 
\end{center}
\end{figure}

In order to investigate portfolio performances within different asset classes, in Figure (\ref{sr_classes}) we show the differences in Sharpe ratio across equities, bonds, commodities and currencies (FX). First, it is interesting to notice the diversification effect when we allow to combine different asset classes in the same portfolio, because none of the individual asset classes in  Figure (\ref{sr_classes}) was able to deliver a Sharpe Ratio as high as in the whole portfolio in Table (\ref{economic_1}). Our dynamic binary classifier method performed specially better among commodities, an asset class traditionally explored in CTAs by the use of trend-following strategies. Regardless of the model specification, applying our econometric approach to commodities delivered substantial improvements compared to the naive benchmark, with even stronger results for TVP-DMS. There was also a small improvement among bonds and similar performance compared to the naive benchmark for equities. The only asset class where our econometric approach clearly performed worse than the benchmark was for the FX class.

\begin{table}[!h] \centering 
  \caption{Economic Performance of TSMOM strategies ($c$ from $CV_{1}$)} 
  \label{economic_2} 
\begin{tabular}{@{\extracolsep{5pt}}l ccccccc} 
\\[-1.8ex]\hline 
\hline \\[-1.8ex] 
 & Turnover & Mean & Vol. & Max.DD & SR & $\Phi$ \\ 
\hline \\[-1.8ex] 
&  &  &  &  &  & \\ 
&  &  & \textbf{CP} &  &  & \\ 
  \hline \\[-1.8ex] 
DMA & $65.9$ & $9.9$ & $10.0$ & $20.1$ & $0.99$ & $179.4$ \\ 
DMS & $79.1$ & $11.6$ & $10.0$ & $17.8$ & $1.16$ & $349.3$ \\ 
1m & $61$ & $8$ & $10.0$ & $26.1$ & $0.80$ & $-16.5$ \\ 
2m & $50.8$ & $6.4$ & $10.0$ & $23.7$ & $0.64$ & $-168.3$ \\ 
4m & $49.4$ & $8$ & $10.0$ & $21.5$ & $0.80$ & $-14.6$ \\ 
6m & $44.7$ & $8.1$ & $10.0$ & $21.7$ & $0.81$ & $-7.3$ \\ 
8m & $43.7$ & $8.5$ & $10.0$ & $20.4$ & $0.85$ & $32.9$ \\ 
10m & $43$ & $8.9$ & $10.0$ & $18.2$ & $0.89$ & $74.5$ \\ 
12m & $44.4$ & $6$ & $10.0$ & $19.1$ & $0.60$ & $-214.4$ \\ 
\hline \\[-1.8ex] 
&  &  &  &  &  & \\ 
&  &  & \textbf{TVP} &  &  & \\ 
  \hline \\[-1.8ex] 
DMA & $84$ & $10.3$ & $10.0$ & $14$ & $1.03$ & $220.5$ \\ 
DMS & $97$ & $12.4$ & $10.0$ & $14.6$ & $1.24$ & $439.2$ \\ 
1m & $75.5$ & $8.3$ & $10.0$ & $16.1$ & $0.83$ & $20.5$ \\ 
2m & $62.5$ & $6.1$ & $10.0$ & $27.1$ & $0.61$ & $-203.6$ \\ 
4m & $55.7$ & $6.5$ & $10.0$ & $28.1$ & $0.65$ & $-165.2$ \\ 
6m & $48.3$ & $8$ & $10.0$ & $19$ & $0.80$ & $-12.3$ \\ 
8m & $51.8$ & $8.2$ & $10.0$ & $17.6$ & $0.82$ & $6.8$ \\ 
10m & $50.6$ & $7.4$ & $10.0$ & $17.7$ & $0.74$ & $-69.8$ \\ 
12m & $51.5$ & $6.3$ & $10.0$ & $18.4$ & $0.63$ & $-177.8$ \\ 
\hline \\[-1.8ex] 

\end{tabular} 
\begin{tablenotes}
      \small 
  \item      \textit{The table reports economic performance from classifier models using constant parameters (CP) and time-varying parameters (TVP), considering single predictors (1m, 2m, 4m, ..., 12m) or applying DMA or DMS with all predictors in the model space. Classifier cutoffs are obtained from a sequential cross-validation procedure over time for each individual asset, where c $\in \{0.49,0.491,...,0.509,0.51\}$. All strategies are scaled to an ex-post annualized volatility of 10\%.}
    \end{tablenotes}
\end{table}

In Tables (\ref{economic_2}) and (\ref{economic_3}) we show results when the classifier cutoffs are obtained from cross-validation procedures, as described in Section (\ref{predictability}). Out-of-sample portfolio performances are still robust for different cutoffs. Model selection and averaging continue to improve final performance outcomes not just in terms of Sharpe Ratios but also in terms of utility gains for the investor. For $CV_{1}$ in Table (\ref{economic_2}), DMS-TVP was able even to slightly improve compare to the fixed $c=50\%$ case. A mean-variance investor would pay 439 bps to give up the naive benchmark strategy to the DMS-TVP in this setting, while she would pay 349 bps for its CP counterpart. For the CP panel the performance is still similar, but with tiny decreases, while for TVP there was small improvements in the overall evaluation.

\begin{table}[!htbp] \centering 
  \caption{Economic Performance of TSMOM strategies ($c$ from $CV_{2}$)} 
  \label{economic_3} 
\begin{tabular}{@{\extracolsep{5pt}}l ccccccc} 
\\[-1.8ex]\hline 
\hline \\[-1.8ex] 
 & Turnover & Mean & Vol. & Max.DD & SR & $\Phi$ \\ 
\hline \\[-1.8ex] 
&  &  &  &  &  & \\ 
&  &  & \textbf{CP} &  &  & \\ 
  \hline \\[-1.8ex] 
DMA & $59.2$ & $9.5$ & $10.0$ & $22.4$ & $0.95$ & $138.1$ \\ 
DMS & $68.2$ & $10.5$ & $10.0$ & $18.2$ & $1.05$ & $239.8$ \\ 
1m & $49.3$ & $7.8$ & $10.0$ & $28.8$ & $0.78$ & $-35.9$ \\ 
2m & $44.5$ & $7.7$ & $10.0$ & $25.1$ & $0.77$ & $-44.9$ \\ 
4m & $44.4$ & $8.1$ & $10.0$ & $23.6$ & $0.81$ & $-2.9$ \\ 
6m & $39.5$ & $7.7$ & $10.0$ & $24.1$ & $0.77$ & $-39.2$ \\ 
8m & $41.6$ & $8.8$ & $10.0$ & $17.9$ & $0.88$ & $70.0$ \\ 
10m & $44.2$ & $8.2$ & $10.0$ & $16.6$ & $0.82$ & $11.4$ \\ 
12m & $38.5$ & $6.3$ & $10.0$ & $20.8$ & $0.63$ & $-177.8$ \\ 
\hline \\[-1.8ex] 
&  &  &  &  &  & \\ 
&  &  & \textbf{TVP} &  &  & \\ 
  \hline \\[-1.8ex] 
DMA & $67.5$ & $9.3$ & $10.0$ & $18.7$ & $0.93$ & $121.6$ \\ 
DMS & $77.2$ & $10.9$ & $10.0$ & $17.9$ & $1.09$ & $277.9$ \\ 
1m & $57.1$ & $8.5$ & $10.0$ & $22.7$ & $0.85$ & $40.0$ \\ 
2m & $50$ & $7.2$ & $10.0$ & $26.4$ & $0.72$ & $-96.5$ \\ 
4m & $44.6$ & $7.4$ & $10.0$ & $23.2$ & $0.74$ & $-69.6$ \\ 
6m & $42$ & $8.3$ & $10.0$ & $19.9$ & $0.83$ & $21.9$ \\ 
8m & $44$ & $8.1$ & $10.0$ & $19.9$ & $0.81$ & $-2.3$ \\ 
10m & $45.1$ & $7.4$ & $10.0$ & $18.1$ & $0.74$ & $-70.3$ \\ 
12m & $42.3$ & $6.6$ & $10.0$ & $15.9$ & $0.66$ & $-149.7$ \\ 
 
\hline \\[-1.8ex] 
\end{tabular} 
\begin{tablenotes}
      \small 
  \item      \textit{The table reports economic performance from classifier models using constant parameters (CP) and time-varying parameters (TVP), considering single predictors (1m, 2m, 4m, ..., 12m) or applying DMA or DMS with all predictors in the model space. Classifier cutoffs are obtained from a sequential cross-validation procedure over time for each individual asset, where c $\in \{0.45,0.46,...,0.54,0.55\}$. All strategies are scaled to an ex-post annualized volatility of 10\%.}
    \end{tablenotes}
\end{table}

For the second cross-validation procedure ($CV_{2}$) in Table (\ref{economic_3}), performances are still strong compared to the naive time-series momentum strategy. There are small portfolio outcomes declines related to those observed when $c=50\%$, but results are still consistent, with DMS and DMA delivering important improvements and TVP outperforming its CP counterpart. It is interesting to notice smaller portfolio turnovers than previous results. Table (\ref{economic_3}) shows that a mean-variance investor would pay 278 bps to give up the naive benchmark strategy to the DMS-TVP, while she would pay 240 bps for its CP counterpart. The DMA setting also performed well, with SR higher than 0.90 for both CP and TVP. Finally, model selection and combination were able to reduce maximum drawdowns, regardless of the dynamics on coefficients.

The results in this section confirm that the discretionary way of building trading positions based solely on the observation of past momentum is not enough to distinguish between future uptrends or downtrends. By ignoring the time-varying patterns of different momentum speeds and its relations with future returns, the investor is giving up the opportunity to sequentially learn about trend instabilities to improve trading signals. Also, although we observe just small economic gains on introducing dynamics on coefficients compared to CP models, it enabled the investor to learn the time-varying relations between momentum speeds and future returns and, as we show in the next section, this time-varying pattern was highlighted during the 2009 TSMOM crash. Therefore, we argue here that by the use of a dynamic classifier model and momentum speed learning, the investor is benefited not just in terms of higher Sharpe Ratio and returns, but also by an increase in final utilities after accounting for a risk aversion measure and considerable reduction on portfolio drawdowns, reducing losses during bad periods for time series momentum strategies.

\section{The 2009 Crash and subsequent periods}
\label{crash_2009}

As we previously discussed, the dynamic classifier was able to dramatically reduce drawdowns. In this section we explore the benefits of automatically let the model learn a turning point indicating a market rebound. It is well known the failure of momentum strategies during market rebounds, in particular the 2009's. \cite{daniel2016momentum} investigates cross-sectional momentum crashes and show that the 2009's was largely impacted by the market rebound. After a period of negative trends, at March of that year the market started to strongly recover. However, at that time, the 12 months momentum strategy was mainly selling assets with high betas (strong positive correlation with the market) and buying assets with low betas (strong negative correlation with the market). As soon as the market recovered, the momentum strategy faced huge losses, since it was selling assets with strong positive recovery and buying assets with weak recovery or even negative growth. 

In the present study we investigate a similar pattern for the naive time-series momentum strategy. During March 2009, the naive benchmark started to suffer huge losses that lasted until June 2010, accumulating a total loss of 25.3\%, representing its maximum drawdown in the last four decades. The literature has already recognized the weakness of TSMOM strategies after the Great Recession (\citealp{baltas2020demystifying}, \citealp{garg2020breaking} and \citealp{garg2021momentum}), where the usual explanation goes from higher asset correlations to the increase of trend breaks but, at the best of our knowlegde, the literature have not discussed the special period of the 2009 market rebound. This important drawback from naive time-series momentum strategies and the lack of academic discussion on the subject motivate us to explore this problematic period and to show how our dynamic classifier approach is able to deal with trend breaks with great sucess.

The great weakness of time-series and cross-sectional momentum strategies is its lack of ability on recognizing such turning points. A longer momentum speed strategy is not able to recognize a sudden trend break, so the investor keeps  following a trend that no longer exists. At the other hand, a very short momentum speed is able to identify new trends when they start to appear, but fails to enter in more stable medium/longer trends. Additionally, sticking to a short momentum speed strategy tend to be less profitable and riskier. Therefore, the great challange is to recognize the periods when this new trends begin by considering informations from fast momentum signals or learning when older long trends disappear. However, we argue here that those patterns are not clear from the simple observation of past returns. Hence, the investor can rely on an econometric approach that is able to digest those patterns in the data. By the use of a dynamic model, the time-varying relations between different momentum speeds and return signals can be capture. Therefore, our dynamic binary classifier approach suits quite well to this kind of decision making problem, since by the use of model selection or model averaging we are able to assign dynamic probabilities for different iterations of momentum speeds, moving from an old type of trend to a new one if it is empirically desirable.

\begin{table}[!htbp] \centering 
  \caption{TSMOM Crash: 2009m03 - 2010m06} 
  \label{crash} 
\begin{tabular}{@{\extracolsep{5pt}}l ccccccc} 
\\[-1.8ex]\hline 
\hline \\[-1.8ex] 
 & Turnover & Accumul. & Mean & Vol. & SR & $\Phi$ \\ 
\hline \\[-1.8ex] 
&  &  &  &  &  & \\ 
&  &  &\textbf{CP}  &  &  & \\ 
  \hline \\[-1.8ex] 
DMA & $64.3$ & $9$ & $6.8$ & $9.7$ & $0.70$ & $3,582$ \\ 
DMS & $73.9$ & $16$ & $11.6$ & $10.3$ & $1.12$ & $4,159$ \\ 
1m & $67.1$ & $31.5$ & $21.2$ & $11.5$ & $1.85$ & $5,396$ \\ 
2m & $50.1$ & $30.6$ & $20.7$ & $11.8$ & $1.75$ & $5,278$ \\ 
4m & $46.1$ & $23.7$ & $16.5$ & $10.4$ & $1.59$ & $4,842$ \\ 
6m & $38.8$ & $14.4$ & $10.3$ & $8.2$ & $1.27$ & $4,207$ \\ 
8m & $47.1$ & $9.4$ & $7.0$ & $8.8$ & $0.79$ & $3,688$ \\ 
10m & $48.1$ & $3.4$ & $2.9$ & $10.6$ & $0.28$ & $2,989$ \\ 
12m & $44.5$ & $-4.7$ & $-3.1$ & $11.1$ & $-0.28$ & $2,187$ \\ 
\hline \\[-1.8ex] 
&  &  &  &  &  & \\ 
&  &  &  \textbf{TVP}&  &  & \\ 
  \hline \\[-1.8ex] 
DMA & $94$ & $15.4$ & $11.0$ & $9.0$ & $1.22$ & $4,222$ \\ 
DMS & $92.3$ & $22.1$ & $15.6$ & $11.8$ & $1.32$ & $4,541$ \\ 
1m & $87.9$ & $41.3$ & $26.9$ & $13.2$ & $2.04$ & $6,003$ \\ 
2m & $60$ & $27.5$ & $19.1$ & $13.4$ & $1.42$ & $4,803$ \\ 
4m & $51.6$ & $30.5$ & $20.6$ & $11.2$ & $1.84$ & $5,346$ \\ 
6m & $56.2$ & $10.7$ & $8.0$ & $10.5$ & $0.76$ & $3,660$ \\ 
8m & $53.1$ & $6.2$ & $4.8$ & $9.7$ & $0.50$ & $3,320$ \\ 
10m & $63.2$ & $0.4$ & $0.7$ & $9.8$ & $0.07$ & $2,781$ \\ 
12m & $45.7$ & $-2.4$ & $-1.4$ & $10.6$ & $-0.13$ & $2,441$ \\
\hline \\[-1.8ex] 
&  &  &  &  &  & \\ 
&  &  & \textbf{Naive} &  &  & \\ 
  \hline \\[-1.8ex] 
1m & $249.7$ & $6.3$ & $4.9$ & $10.0$ & $0.49$ & $3,305$ \\ 
2m & $180.9$ & $11.2$ & $8.7$ & $13.6$ & $0.64$ & $3,369$ \\ 
4m & $114.7$ & $11$ & $8.2$ & $9.9$ & $0.83$ & $3,752$ \\ 
6m & $84.6$ & $1.9$ & $1.9$ & $11.2$ & $0.17$ & $2,791$ \\ 
8m & $73.3$ & $-6.3$ & $-4.1$ & $13.6$ & $-0.30$ & $1,778$ \\ 
10m & $59.5$ & $-14.9$ & $-11.3$ & $13.6$ & $-0.83$ & $976$ \\ 
12m & $72.1$ & $-25.3$ & $-20.9$ & $13.3$ & $-1.57$ & $0.0$ \\ 
\hline \\[-1.8ex] 
\end{tabular} 
\begin{tablenotes}
      \small 
   \item     \textit{The table reports economic performance during the time-series momentum crash period (2009m03 until 2010m06) from classifier models using CP and TVP, considering single predictors (1m, 2m, 4m, ..., 12m) or applying DMA or DMS with all predictors in the model space. Results consider a classifier cutoff of $c=50\%$.}
    \end{tablenotes}
\end{table}

In Table (\ref{crash}) we compare the performance of the dynamic classifier method and the Naive-TSMOM during the crash period. The second column of the table shows the accumulated return during the crash, where the Naive-TSMOM benchmark suffered 25.3\% of losses. In fact, one aspect that is not well discussed in the literature is that the naive benchmark was able to recover from those losses just at the end of 2014! The 8 and 10-month naive strategies also delivered negative returns during the crash period, while shorter momentum straties were able to survive the crash period with positive returns. Interesting, the 4-month naive strategy presented a Sharp ratio of 0.83, while the 12-month benchmark had a strong negative return adjusted by risk of -1.57. 

Table (\ref{crash}) gives evidences of failure in traditional longer TSMOM signals to antecipate drastic trend changes. At the other hand, by looking for the first and second panels of the table, we notice that our dynamic binary classifier was able to perform extremely better than the naive approach. As it was expected, fast momentum predictors delivered a very high accumulated returns, in speciall when TVP are allowed. The 1-month single predictor for the TVP setting obtained 41.3\% of total accumulated excess returns during the crash period with a impressive Sharpe Ratio of 2.04. The only look-back period delivering a negative performance was the 12-month single predictor, but the losses were tiny compared to the naive approach. However, as mentioned before, sticking solely to a very fast look-back period can induce lower performances in the long-run, then recognizing the time-varying importance of different momentum speeds is crucial for a stronger portfolio with lower risk and drawdowns. Since the DMA and DMS settings were built exactly with the goal of learning different momentum speed dynamics, we can notice that they provide strong portfolio performances not just on the long-run as we have shown in last sections, but also during the 2009 momentum crash period. Table (\ref{crash}) also makes clear the advantage of allowing time-varying parameters, since during bad periods the economic relations among financial data can change in a matter of few periods. When the investor considers the DMS-TVP approach, she is able to obtain 22.1\% of accumulated returns during the momentum crash with a robust SR of 1.32! It means that a mean-variance investor would pay 4,541 bps to switch from the 12-month naive approach to the DMS-TVP method.

\begin{figure}[!h]
\begin{center}
\caption{Mean inclusion probabilities - Momentum speeds}
\label{l_probs}
\includegraphics[width=0.8\textwidth]{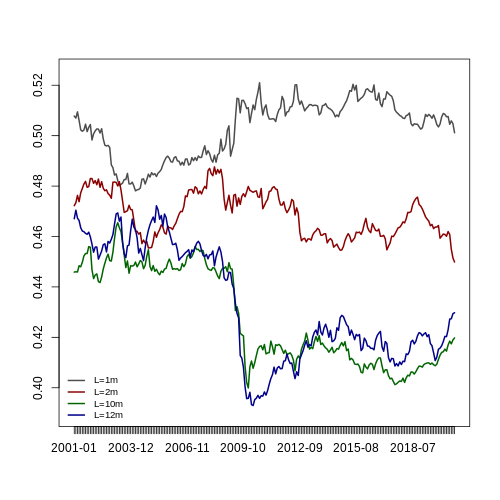}
 
\end{center}
\end{figure}

In order to provide evidences that the dynamic classifier method with TVP is capable of learning from past mistakes and assign higher probabilities for those look-back periods that are performing better in the recent past and reducing probabilities for trends that no longer exists, Figure (\ref{l_probs}) shows inclusion probabilities for momentum predictors of 1, 2, 10 and 12-months, averaged across all different assets. For a given asset, a inclusion probabilitity (IP) for a specific momentum speed $L$ can be defined as

$$
IP^{L} = \sum_{j=1}^{M} \mathbb{1}_{(j \subset J)} p\left(M_{j} | \mathcal{D}_{t}\right)
$$

\noindent where $J$ represents the subset of models containing the specific momentum predictor $L$ and $\mathbb{1}_{(j \subset J)}$ is an indicator function taking the value of 1 if the model $j$ is cointained on J. Hence, a higher $IP^{L}$ means that models with the momentum predictor $L$ are performing better in the recent past and then receiving higher model probabilities. Since we average $IP^{L}$ for all assets available at the time period, Figure (\ref{l_probs}) can give us a sense of the overall importance of longer ou shorter trends over time and in special the 2009 momentum Crash. 

 It is evident that as soon as the rebound starts at the beginning of 2009, models with look-back period of 10 and 12-month momentum predictors sequentially received lower probabilities while the 1-month momentum predictor increase in importance. The 2-month momentum predictor continued to oscillate around its older inclusion probabilities, but remaining higher than longer momentum predictors. It is interesting to notice that since the 2009 Crash, longer momentum speeds remained much less important than faster momentum speeds, in line with recent evidences on the increase of trend breaks since 2010 (\citealp{baltas2020demystifying}, \citealp{garg2020breaking} and \citealp{garg2021momentum} ). At the same time, the 1-month look-back period remains as the predictor with highest inclusion probabilities. In fact, although longer trends were more important than they are nowadays, the 1-month momentum already had greater importance even before the Great Recession. Therefore, Figure (\ref{l_probs}) gives evidences that sequentially learning the importance of each momentum speed, combining or selecting those different informations to generate out-of-sample signal forecasts was able to deal with the 2009 trend break problem with great sucess, as Table (\ref{crash}) highlights.

\begin{table}[!htbp] \centering 
  \caption{Economic Performance: 2010m07 - 2020m09} 
  \label{2010_2020} 
\begin{tabular}{@{\extracolsep{5pt}}l ccccccc} 
\\[-1.8ex]\hline 
\hline \\[-1.8ex] 
 & Turnover & Mean & Vol. & Max.DD & SR & $\Phi$ \\ 
\hline \\[-1.8ex] 
&  &  &  &  &  & \\ 
&  &  & \textbf{CP} &  &  & \\ 
  \hline \\[-1.8ex] 
DMA & $65$ & $9.7$ & $10.0$ & $14.8$ & $0.97$ & $470.6$ \\ 
DMS & $80.4$ & $11.8$ & $10.0$ & $10.4$ & $1.18$ & $698.9$ \\ 
1m & $64.7$ & $11.8$ & $10.0$ & $10.1$ & $1.18$ & $692.9$ \\ 
2m & $47.9$ & $9.8$ & $10.0$ & $11.2$ & $0.98$ & $484.5$ \\ 
4m & $47.4$ & $7.0$ & $10.0$ & $18.9$ & $0.70$ & $197.2$ \\ 
6m & $41.7$ & $8.1$ & $10.0$ & $14.9$ & $0.81$ & $311.1$ \\ 
8m & $38.3$ & $9.0$ & $10.0$ & $12.5$ & $0.90$ & $407.5$ \\ 
10m & $35.7$ & $9.0$ & $10.0$ & $16.8$ & $0.90$ & $406.3$ \\ 
12m & $35.2$ & $8.7$ & $10.0$ & $15.5$ & $0.87$ & $372.7$ \\ 
\hline \\[-1.8ex] 
&  &  &  &  &  & \\ 
&  &  & \textbf{TVP} &  &  & \\ 
  \hline \\[-1.8ex] 

DMA & $93.4$ & $9.4$ & $10.0$ & $12.8$ & $0.94$ & $440.1$ \\ 
DMS & $114.5$ & $11.0$ & $10.0$ & $13.7$ & $1.10$ & $609.3$ \\ 
1m & $82.7$ & $9.3$ & $10.0$ & $11.1$ & $0.93$ & $433.5$ \\ 
2m & $60.5$ & $7.2$ & $10.0$ & $14$ & $0.72$ & $222.6$ \\ 
4m & $54.2$ & $7.4$ & $10.0$ & $17$ & $0.74$ & $233.5$ \\ 
6m& $48.1$ & $8.4$ & $10.0$ & $13.5$ & $0.84$ & $338.2$ \\ 
8m & $45.5$ & $7.5$ & $10.0$ & $17.2$ & $0.75$ & $246.8$ \\ 
10m & $39.0$ & $6.3$ & $10.0$ & $21.2$ & $0.63$ & $128.3$ \\ 
12m & $48.4$ & $6.1$ & $10.0$ & $21.7$ & $0.61$ & $108.2$ \\ 
\hline \\[-1.8ex] 
&  &  &  &  &  & \\ 
&  &  & \textbf{Naive} &  &  & \\ 
  \hline \\[-1.8ex] 
1m & $356.8$ & $0.8$ & $10.0$ & $27$ & $0.08$ & $-412.7$ \\ 
2m & $225.3$ & $5.4$ & $10.0$ & $13$ & $0.54$ & $31.6$ \\ 
4m & $163.8$ & $2.9$ & $10.0$ & $19.5$ & $0.29$ & $-216.0$ \\ 
6m & $127.7$ & $2.8$ & $10.0$ & $17.4$ & $0.28$ & $-219.2$ \\ 
8m & $118.5$ & $2.4$ & $10.0$ & $13.5$ & $0.24$ & $-265.8$ \\ 
10m & $95.1$ & $5.4$ & $10.0$ & $16.3$ & $0.54$ & $38.8$ \\ 
12m & $89.1$ & $5.0$ & $10.0$ & $15.5$ & $0.50$ & $0.0$ \\ 
\hline \\[-1.8ex] 
\end{tabular} 
\begin{tablenotes}
      \small 
    \item    \textit{The table reports economic performance after the time-series momentum crash period (2010m07 until 2020m09) from classifier models using CP and TVP, considering single predictors (1m, 2m, 4m, ..., 12m) or applying DMA or DMS with all predictors in the model space. Results consider a classifier cutoff of $c=50\%$. All strategies are scaled to an ex-post annualized volatility of 10\%.}
    \end{tablenotes}
\end{table}

\subsection{The post-Crash period}

After the Great Recession, the number of turning points increased considerably for different assets.  \cite{garg2020breaking} show that there is a negative relation among the number of turning points and Sharpe Ratios of TSMOM strategies. In order to show the robustness of our dynamic classifier approach on the subsequent periods of the 2009 Crash, Table (\ref{2010_2020})  displays portfolio results from July 2010 to September 2020. 

Our results confirm the weakness of the Naive-TSMOM strategies on the post Great Financial Crisis, as observed in the works of \cite{rubesam2020long}, \cite{garg2020breaking} and \cite{baltas2020demystifying}. The naive benchmark strategy obtained a Sharpe ratio of 0.50 during the period, which means a reduction of about 40\% compared to the whole sample evaluated before and the results remain weaker regardless of the look-back period considered.

When our binary classifier is applied, the optics is still very optimistic. Indeed, what can be seen is a much stronger performance for the subsequent period of the 2009 Crash compared to the Naive-TSMOM. For any single predictor setting, the performances are better than the naive benchmark. Portfolio improvements are observed regardless of the dynamics induced in coefficients. The DMS-TVP was able to generate a Sharpe Ratio of 1.10, a 120\% increase in relation to the naive benchmark, which would require an annualized management fee of 609.3 bps for the investor give up the traditional naive benchmark to start using the DMS-TVP model. It is interesting to notice that, for this particular sample period, the CP setting performed even better than the model settings where TVP are allowed. The DMS-CP have showed a Sharpe ratio of 1.18, representing 698.9 bps as management fees to use switch from the naive benchmark to this particular model setting. These results demonstrate evidence of no incremental performance for dynamics on coefficients and a simple constant parameter binary classifier model is able to successfully deal with the amount of turning points in the post 2009 Crash. The most important pattern observed in the period is the strong performance of the single 1-month momentum predictor. The results are in line with Figure (\ref{l_probs}), where the 1-month look-back emerged with higher inclusion probabilities than longer/slower momentum measures.

Finally, just for the sake of curiosity, one of the highest drawdowns from the naive benchmark strategy was exactly during the Covid period. Since April 2020 to September of that year, the traditional 12-month time-series momentum strategy accumulated 8.6\% of return losses. The performance  was not worse in 2020 because many assets at the very beginning of the year were signaling negative momentum such that, when the market really suffered huge losses in March, the strategy was able to profit from negative trends, earning 8.7\% in that month. Hence, from March to Semptember, the naive benchmark accumulated just 0.6\% of losses. At the other hand, the DMS-TVP was able to deliver 18.4\% of accumulated returns from March to September. Its CP counterpart, the DMS-CP model, also performed quite well in this period, earning 15.0\% of accumulated returns.

Therefore, Table (\ref{2010_2020}) gives evidences that the the dynamic classifier performance is robust even for periods of higher trend breaks. Since dynamic model probabilities are able to dynamically assign higher or lower probabilities to models with different momentum speeds, as we have showed in Section \ref{econometric}, new financial environments are not enough to weaken its final portfolio performance. What we actually see is the oposite, where period of higher changes in financial trends are accompanied by better returns adjusted for risk.

\section{Conclusion}
\label{Conclusion}

Since the work of \cite{moskowitz2012time}, the literature on trend-following strategies has grown rapidly and its applicability has spread throughout the financial industry. However, there is still a lack of discussion of how to incorporate econometric models to help investors to learn about better momentum speeds over time. From the investor perspective, the better understanding of the time-varying relations from past accumulated returns and future return signals are crucial for portfolio construction. Recent evidences has shown that standard discretionary time series strategies tend to suffer stronger breaking trends and crashes, which dramatically harm portfolio returns adjusted for risk. 

In this study we propose the use of a dynamic binary classifier model where investors can sequentially learn the sensitivities between past  returns and future signals. Imposing time-varying parameters, the model is able to adapt to changes in the financial market, moving faster from momentum to reversal if it is empirically wanted. Also, by the use of dynamic model probabilities, the approach is able to recognize sudden turning points, sequentially switching from slow to fast momentums after a market rebound, dramatically reducing drawdowns and momentum crashes. Our results show not just better forecasting accuracy gains compared to the naive time series momentum strategy but also that an investor using the dynamic classifier approach earns annualized Sharpe Ratios much higher than the naive benchmark. We analyze different model specifications, cutoffs and subsamples and results still have shown robustness. The performances remained quite strong even after the Great Financial Crisis. Considering a mean-variance investor with a quadratic utility, we show that she will be willing to pay an annualized management fee of 425.1 basis points to switch from the naive 12 months time series momentum strategy to our dynamic classifier approach with model selection and time-varying coefficients. Therefore, it generates not just strong portfolio performance, but great economic utility gains for investors. We show that utility gains are even higher during the 2009 momentum Crash and in the last decade.  Those are good news for portfolio managers who are interested in improving trend investment strategies in a unstable financial world with high model uncertainties and rapid and complex changes over time.

The strong results obtained using future contracts, in special among commodity futures, motivate us to consider as an extension for future research the use of a larger set of commodities to be analyzed. Also, we also pretend to extend to a larger cross-section of equity returns. Inspired by the recent works of \cite*{jiang2020re} and \cite*{kelly2021understanding}, where the authors also apply econometric forecasting models to portfolio construction, our future interest is to test the dynamic classifier approach for cross-sectional momentum strategies, building long-short portfolios by different quantiles of model ranking predictions. We believe that this extension can be seen as a strong return forecasting model competitor for the recent advances in the momentum literature.

%##########################

\clearpage

\appendix
\setcounter{secnumdepth}{0}

\section{Appendix: Aditional Results}
\label{appendix}

\subsection{Bayesian portfolio decision}

We explain here the implicit cutoff selection obtained from an Bayesian decision perspective. The main goal of the Bayesian investor is to sequentially select the action  that maximizes expected utility. For each period of time $t$ and for each asset available $i$, the investor is faced with two simple actions within the set of possible actions $\mathcal{A} = \{Long, Short \}$, i.e., she can open a long or short position for asset $i$. After the realization of the true return value, each given action can produce a different utility for the investor and this utility will depend on the actual return direction for that period $t$. If the investor went long an asset and after observing its true direction it was actually up (positive), then the investor should receive a positive utility gain. The same would apply if she went short an asset that was actually down (negative). When the action made by the investor does not match the actual direction, she should lose utility. The Table below summarizes the possible actions and their final outcomes.

\vspace{1cm}

\begin{tabular}{l|l|c|c|c}
\multicolumn{2}{c}{}&\multicolumn{2}{c}{Actual Directions}&\\
\cline{3-4}
\multicolumn{2}{c|}{}&Positive&Negative&\multicolumn{1}{c}{}\\
\cline{2-4}

Actions ($\mathcal{A}$)& Long & $U_{L,P}$ & $U_{L,N}$ & \\
\cline{2-4}
& Short & $U_{S,P}$ & $U_{S,N}$ & \\
\cline{2-4}
\end{tabular}

\vspace{1cm}

\noindent where 

\begin{itemize}
    \item $U_{L,P}$ is the utility when a Long position is opened and the actual return was Positive;
    
    \item $U_{L,N}$ is the utility when a Long position is opened and the actual return was Negative;
     
    \item $U_{S,P}$ is the utility when a Short position is opened and the actual return was Positive;
    
    \item $U_{S,N}$ is the utility when a Short position is opened and the actual return was Negative

\end{itemize}

Since we consider a quadratic utility for the mean-variance investor in the same spirit of Equation (\ref{eq_utility}), we compute:

\begin{itemize}
    \item $U_{L,P} = \bar{R}_{(+)} - \frac{\gamma}{2(1+\gamma)} \bar{\sigma}_{(+)}$
    
    \item $U_{L,N} = \bar{R}_{(-)} - \frac{\gamma}{2(1+\gamma)} \bar{\sigma}_{(-)}$ 
     
    \item $U_{S,P} = -\bar{R}_{(+)} - \frac{\gamma}{2(1+\gamma)} \bar{\sigma}_{(+)}$ 
    
    \item $U_{S,N} = -\bar{R}_{(-)} - \frac{\gamma}{2(1+\gamma)} \bar{\sigma}_{(-)}$

\end{itemize} 

\noindent where the bar upscript represents historical sample estimates until the decision period and the signs subscripts in parentheses filter for positive or negative historical observations. Hence, the investor considers those utility estimates as possible final outcomes before assuming a specific action. 

At the end of time $t-1$, the investor will choose the action that maximizes his expected utility for time $t$, where the expected utility for each action will depend on the forecasting output for asset $i$ from the model she is considering. As an example, suppose the investor is willing to open a position on a specific asset $i$ and will consider to use DMA as a forecasting model to decide the probability of a positive return on the next period. Hence, she will use the forecasting output as in Equation (\ref{eq_dma}) to compute expected utility for a Long position ($E[U(Long)]$) and for Short position ($E[U(Short)]$):

$$
E[U(Long)] = \widehat{s}_{t|t-1}^{DMA} U_{L,P} + (1 -  \widehat{s}_{t|t-1}^{DMA}) U_{L,N}
$$

and 

$$
E[U(Short)] = \widehat{s}_{t|t-1}^{DMA} U_{S,P} + (1 -  \widehat{s}_{t|t-1}^{DMA}) U_{S,N}
$$

If $E[U(Long)] \geq E[U(Short)]$, the investor opens a long position on asset $i$, and goes short otherwise.

Table (\ref{expect_utility}) below show results when the investor sequentially applies the mechanism explained above for each asset available across time. It is important to highlight here that the Bayesian decision is applied just to produce trading position signs, but portfolio construction and weighing still follows the same structure as Equations (\ref{dma_port}) and (\ref{dms_port}).

The general conclusions in Table (\ref{expect_utility})  follow the same we have obtained before in the main body of the paper.

\begin{table}[!htbp] \centering 
  \caption{Economic - Expected Utility decision} 
  \label{expect_utility} 
\begin{tabular}{@{\extracolsep{5pt}}l ccccccc} 
\\[-1.8ex]\hline 
\hline \\[-1.8ex] 
 & Turnover & Mean & Vol. & Max.DD & SR & $\Phi$ \\ 
\hline \\[-1.8ex] 
&  &  &  &  &  & \\ 
&  &  & \textbf{CP} &  &  & \\ 
  \hline \\[-1.8ex] 
DMA & $68.2$ & $10.3$ & $10.0$ & $14.5$ & $1.030$ & $214.480$ \\ 
DMS & $79.7$ & $11.2$ & $10.0$ & $14.5$ & $1.120$ & $306.910$ \\ 
1m & $75.4$ & $7.3$ & $10.0$ & $25.3$ & $0.730$ & $-87.470$ \\ 
2m & $59.5$ & $6.2$ & $10.0$ & $23$ & $0.620$ & $-186.910$ \\ 
4m & $52.9$ & $8$ & $10.0$ & $20.4$ & $0.8$ & $-11.260$ \\ 
6m & $45.7$ & $8.1$ & $10.0$ & $14.1$ & $0.810$ & $-7.260$ \\ 
8m & $45.6$ & $8.8$ & $10.0$ & $18.4$ & $0.880$ & $69.840$ \\ 
10m & $43.9$ & $8.8$ & $10.0$ & $17.1$ & $0.880$ & $62.780$ \\ 
12m & $42.6$ & $6.3$ & $10.0$ & $17.5$ & $0.630$ & $-183.630$ \\ 

\hline \\[-1.8ex] 
&  &  &  &  &  & \\ 
&  &  & \textbf{TVP} &  &  & \\ 
  \hline \\[-1.8ex] 
DMA & $78$ & $10.5$ & $10.0$ & $11.8$ & $1.050$ & $238.070$ \\ 
DMS & $92.4$ & $11.1$ & $10.0$ & $12.5$ & $1.110$ & $302.450$ \\ 
1m & $80.1$ & $8.9$ & $10.0$ & $16.5$ & $0.890$ & $82.040$ \\ 
2m & $63$ & $6.3$ & $10.0$ & $25.8$ & $0.630$ & $-177.340$ \\ 
4m & $55.4$ & $6.9$ & $10.0$ & $26.1$ & $0.690$ & $-118.730$ \\ 
6m & $49.6$ & $7.6$ & $10.0$ & $16.7$ & $0.760$ & $-49.180$ \\ 
8m & $47$ & $9.2$ & $10.0$ & $12.9$ & $0.920$ & $106.040$ \\ 
10m & $47.1$ & $8.2$ & $10.0$ & $16.7$ & $0.820$ & $7.380$ \\ 
12m & $47.8$ & $6.4$ & $10.0$ & $15.4$ & $0.640$ & $-175.440$ \\ 
\hline \\[-1.8ex] 
\end{tabular} 
\begin{tablenotes}
      \small 
    \item    \textit{The table reports economic performance from classifier models using constant parameters (CP) and time-varying parameters (TVP), considering single predictors (1m, 2m, 4m, ..., 12m) or applying DMA or DMS with all predictors in the model space. Classifier cutoffs are implicit obtained from a sequential Bayesian decision problem, where the investor selects the best trading action based on its expected utility. All strategies are scaled to an ex-post annualized volatility of 10\%.}
    \end{tablenotes}
\end{table} 

\break 

\subsection{Descriptive Statistics}

\begin{table}[!htbp] \centering
  \caption{Summary Statistics of Futures Contracts}
  \label{summary}
\begin{tabular}{@{\extracolsep{5pt}}l ccccc}
\\[-1.8ex]\hline
\hline \\[-1.8ex]
Futures & Start.Date & Mean & Vol & SR \\
\hline \\[-1.8ex]
 \textbf{Equities} & &  &  &  \\ 
$S\&P$500 & 1982-05 & 8.7 & 15.4 & 0.36 \\
Nasdaq 100& 1999-09 & 7.4 & 24.7 & 0.23 \\
$S\&P$ Canada 60  & 1999-10 & 4.5 & 20.5 & 0.15 \\
FTSE 100 & 1984-06 & 4.6 & 18.1 & 0.1 \\
DAX & 1990-12 & 7 & 22.2 & 0.25 \\
CAC 40 & 1999-02 & 0.7 & 21 & -0.02 \\
IBEX 35 & 1992-05 & 2.1 & 23.8 & 0.03 \\
FTSE MIB  & 2004-04 & -2.3 & 25.3 & -0.08 \\
AEX & 1988-11 & 5.1 & 20.3 & 0.14 \\
SMI & 1990-12 & 7.8 & 16.2 & 0.36 \\
Nikkei 225 & 1990-10 & 0.3 & 21.2 & -0.1 \\
ASX SPI 200  & 2000-06 & 4.2 & 22.3 & 0.17 \\
 & &  &  &  \\ 
  \textbf{Bonds} & &  &  &  \\

US Tsy Note 2-year & 1990-07 & 1.3 & 1.5 & -0.74 \\
US Tsy Note 5-year & 1988-06 & 2.7 & 3.9 & -0.02 \\
US Tsy Bond 10-year & 1982-06 & 4.4 & 6.6 & 0.14 \\
US Tsy Bond 30-year & 1980-01 & 4.6 & 11.1 & 0.05 \\
Euro Schatz 2-year & 1998-11 & 0.7 & 9.6 & -0.07 \\
Euro Bobl 5-year & 1999-02 & 2.4 & 9.7 & 0.11 \\
Euro Bund 10-year & 1998-11 & 3.8 & 10.2 & 0.24 \\
Euro Buxl 30-year & 2005-10 & 6.1 & 12.4 & 0.4 \\
UK Long Gilt & 1982-12 & 2.4 & 12.2 & -0.07 \\
Canadian 10-year & 1989-10 & 3.3 & 9.1 & 0.09 \\
Japanese 10-year & 1987-01 & 3.9 & 12.4 & 0.08 \\
 & &  &  &  \\
   \textbf{Currencies} & &  &  &  \\

AUD/USD & 1983-04 & 0.8 & 36.6 & -0.08 \\
CAD/USD & 2007-08 & -4.8 & 40.1 & -0.12 \\
CHF/USD & 2006-02 & -3.2 & 33.1 & -0.12 \\
EUR/USD & 1990-05 & 0.5 & 51.3 & -0.08 \\
GBP/USD & 2005-11 & -2 & 47.1 & -0.07 \\
JPY/USD & 2007-01 & 0.4 & 27 & 0.02 \\
NZD/USD & 1980-01 & 3.1 & 17.4 & -0.05 \\
SEK/USD  & 1980-01 & 6.4 & 34.5 & 0.08 \\
\hline \\[-1.8ex] 
\end{tabular} 
\end{table}

\begin{table}[!htbp] \centering 
  \caption{Summary Statistics of Futures Contracts (Continued)} 
  \label{summary_2} 

\begin{tabular}{@{\extracolsep{5pt}}l ccccc} 
\\[-1.8ex]\hline 
\hline \\[-1.8ex] 
Futures & Start Date & Mean & Vol. & SR \\ 
\hline \\[-1.8ex] 

  \textbf{Commodities} & &  &  &  \\ 

Light crude oil &  1980-01 & 0.6 & 25.1 & -0.12 \\
Brent crude oil & 1980-01 & -0.9 & 32.7 & -0.15 \\
Heating oil &  1980-01 & 1.3 & 15.3 & -0.18 \\
Natural gas & 1980-02 & 1.2 & 17.8 & -0.15 \\
RBOB gasoline &  1980-01 & 1.4 & 36.3 & -0.04 \\
Copper  & 1980-01 & 0.7 & 27.3 & -0.12 \\
Gold &2006-04 & 3.4 & 36.5 & 0.11 \\
Palladium &2006-04 & 2.6 & 24.1 & 0.07 \\
Platinum &2006-04 & 4.6 & 31.7 & 0.15 \\
Silver & 1980-01 & 1.1 & 24.6 & -0.09 \\
Feeder cattle &  2006-04 & 3.5 & 33.3 & 0.11 \\
Live cattle & 1980-01 & -0.4 & 30 & -0.14 \\
Lean hogs &1980-01 & -1.2 & 35.3 & -0.19 \\
Corn &  1980-01 & -0.3 & 30.5 & -0.15 \\
Oats & 1980-01 & 2.6 & 35.4 & -0.02 \\
Soybean oil & 1980-01 & 0.4 & 31.8 & -0.16 \\
Soybean meal & 1980-01 & -0.5 & 41.3 & -0.08 \\
Soybeans &  1987-02 & 2.7 & 11.4 & 0.01 \\
Wheat & 1980-01 & -0.3 & 8.8 & -0.49 \\
Cocoa & 1980-01 & -0.7 & 11.9 & -0.38 \\
Coffee & 1999-04 & -0.3 & 9.7 & -0.16 \\
Cotton \#2& 1980-01 & -0.5 & 10.2 & -0.42 \\
Lumber & 1980-01 & -1.1 & 11.6 & -0.44 \\
Orange juice &  1997-06 & 2.2 & 12.8 & 0.04 \\
Sugar \#11 & 2004-05 & -1.7 & 11.2 & -0.23 \\
\hline \\[-1.8ex]
\end{tabular}
\end{table}

\clearpage

\bibliographystyle{ecta}
\bibliography{referencias}
\clearpage

\end{document}